\documentclass{elsart}

\usepackage[square,comma]{natbib}
\usepackage{graphicx}
\usepackage{pxfonts}
\usepackage{lineno}

\usepackage{amssymb}

\journal{}

\begin{document}

\thispagestyle{empty}
\begin{Large}
\textbf{DEUTSCHES ELEKTRONEN-SYNCHROTRON}

\textbf{\large{Ein Forschungszentrum der
Helmholtz-Gemeinschaft}\\}
\end{Large}

DESY 10-133

August 2010

\begin{eqnarray}
\nonumber &&\cr \nonumber && \cr \nonumber &&\cr
\end{eqnarray}
\begin{eqnarray}
\nonumber
\end{eqnarray}
\begin{center}
\begin{Large}
\textbf{Cost-effective way to enhance the capabilities of the LCLS baseline}
\end{Large}
\begin{eqnarray}
\nonumber &&\cr \nonumber && \cr
\end{eqnarray}

\begin{large}
Gianluca Geloni,
\end{large}
\textsl{\\European XFEL GmbH, Hamburg}
\begin{large}

Vitali Kocharyan and Evgeni Saldin
\end{large}
\textsl{\\Deutsches Elektronen-Synchrotron DESY, Hamburg}
\begin{eqnarray}
\nonumber
\end{eqnarray}
\begin{eqnarray}
\nonumber
\end{eqnarray}
ISSN 0418-9833
\begin{eqnarray}
\nonumber
\end{eqnarray}
\begin{large}
\textbf{NOTKESTRASSE 85 - 22607 HAMBURG}
\end{large}
\end{center}
\clearpage
\newpage
\begin{frontmatter}




\title{Cost-effective way to enhance the capabilities of the LCLS baseline}


\author[XFEL]{Gianluca Geloni\thanksref{corr},}
\thanks[corr]{Corresponding Author. E-mail address: gianluca.geloni@xfel.eu}
\author[DESY]{Vitali Kocharyan}
\author[DESY]{and Evgeni Saldin}

\address[XFEL]{European XFEL GmbH, Hamburg, Germany}
\address[DESY]{Deutsches Elektronen-Synchrotron (DESY), Hamburg,
Germany}

\begin{abstract}
This paper discusses the potential for enhancing the LCLS hard X-ray FEL capabilities. In the hard X-ray regime, a high longitudinal coherence will be the key to such performance upgrade. The method considered here to obtain high longitudinal coherence is based on a novel single-bunch self-seeding scheme exploiting a single crystal monochromator, which is extremely compact and can be straightforwardly installed in the LCLS baseline undulator. We present simulation results dealing with the LCLS hard X-ray FEL, and show that this method can produce fully-coherent X-ray pulses at  100 GW power level. With the radiation beam monochromatized down to the Fourier transform limit, a variety of very different techniques leading to further improvements of the LCLS performance become feasible. In particular, we describe an efficient way for obtaining full polarization control at the LCLS hard X-ray FEL. We also propose to exploit crystals in the Bragg reflection geometry as movable deflectors for the LCLS X-ray transport systems. The hard X-ray beam can be deflected of an angle of order of a radian without perturbations. The monochromatization of the output radiation constitutes the key for reaching such result. Finally, we  describe
a new optical pump - hard X-ray probe technique which will allow
time-resolved studies at the LCLS baseline on the femtosecond time scale. The principle of operation of the proposed scheme is essentially based on the use of the time jitter between pump and probe pulses. This eliminates the need for timing XFELs to high-power conventional lasers with femtosecond accuracy.

\end{abstract}

%
%
%
\end{frontmatter}



\section{\label{sec:intro} Introduction}

The LCLS, the world's first hard X-ray FEL, has demonstrated SASE lasing and saturation at 0.15 nm \cite{LCLS2}. The SASE X-ray beams are characterized by full transverse coherence, but due to the inherent nature of the SASE process, the pulses only possess partial longitudinal coherence. This is a consequence of the fact that in SASE FEL the amplification process start up from noise. The coherence time is defined by inverse spectral width.

For conventional XFELs \cite{LCLS2}-\cite{SPRIN}, the typical pulse bandwidth is much larger than the Fourier transform limited value for the radiation pulse duration.

Self-seeding schemes have been studied to reduce the bandwidth of SASE X-ray FELs \cite{SELF}-\cite{OURY3}. A self-seeded XFEL consists of two undulators with an X-ray monochromator located between them. The first self-seeding scheme was proposed in the VUV and soft X-ray region, and makes use of a grating monochromator \cite{SELF}. The implementation of such scheme for hard X-ray range would exploit a  crystal monochromator \cite{SXFE}. Self-seeding schemes proposed in \cite{SXFE} makes use of a four-crystal, fixed-exit monochromator in Bragg reflection geometry. The X-ray pulse thus acquires a cm-long path delay, which must be compensated. For a single-bunch self-seeding scheme this requires a long electron beam bypass, implying modifications of the baseline undulator configuration. To avoid this problem, a double bunch self-seeding scheme based on special photoinjector setup was proposed in \cite{OURL,HUAN}, based on the double-bunch production scheme \cite{DOUB}.

In \cite{OURX}-\cite{OURY3} we proposed to use a new method of monochromatization exploiting a single crystal in Bragg transmission geometry. A great advantage of this method is that the single crystal monochromator introduces no path delay of X-rays. This fact eliminates the need for long electron beam bypass, or for the creation of two precisely separated identical electron bunches, as required in the other self-seeding schemes. In some experimental situations, the simplest two-undulator configuration for self-seeding scheme is not optimal. The obvious and technically possible extension is to use a setup with three or more undulators separated by monochromators. This cascade scheme, exemplified in \cite{OURY2} for the baseline undulators at the European XFEL, is distinguished, in performance, by a small heat-loading of crystals at a very high repetition rate.

The situation with heat loading of crystal changes when one considers the case of LCLS, which is characterized by much lower repetition rate. In this case, the simplest two-undulator configuration may be practically advantageous. This paper describes such almost trivial setup for the LCLS baseline. Using  a single crystal installed within a short ($4$ m-long) magnetic chicane in the LCLS baseline undulator, it is possible to decrease the bandwidth of the radiation well beyond the LCLS SASE FEL design, down to the Fourier limit given by the finite duration of the pulse. In the single-bunch self-seeding scheme proposed here for LCLS, no time dependent elements are used, and problems with synchronization and time jitter do not exist at all. The installation of the chicane does not perturb the undulator focusing system and allows for a safe return to the baseline mode of operation. The inclusion of a chicane is not expensive and may find many other applications \cite{OUR01}-\cite{OUR05}. We present simulation results for the LCLS baseline undulator, with electron beam parameters based on the LCLS commissioning results. In particular, we will use the LCLS low-charge mode of operation ($0.02$ nC) to demonstrate the feasibility of proposed scheme. The wakefield effects from the linac and from the undulator vacuum chamber are much reduced at such low charge.

At that point, there still remains room for improving the output characteristics of the XFEL. The most promising way to increase the output power is by tapering the magnetic field of the undulator \cite{TAP1}-\cite{TAP4}. Significant further increase in power is achievable by starting the FEL process from a monochromatic seed rather than from noise \cite{FAWL,CORN,WANG}.  The baseline LCLS undulator is long enough (33 cells) to reach a 100 GW level monochromatic output by tapering the undulator parameter of about $1 \%$  over the last 10 cells, following monochromatization (11 cells) and saturation (12 cells). It should be further noted that the proposed self-seeding scheme could be applied at higher charge modes of operation as well. In particular, the $0.25$ nC mode of operation of the LCLS, relying on the self-seeded hard X-ray technique, ultimately leads to an increase in peak brilliance of two orders of magnitude.

Once highly monochromatic X-ray beam can be obtained from the baseline undulator, a variety of very different techniques for further performance improvement of the LCLS  become feasible. In particular, the high monochromatization of the output radiation constitutes the key for obtaining full polarization control
at the LCLS hard X-ray FELs. So far, two  different approaches
have been used for the production of circularly-polarized photons
with synchrotron radiation; (i) special insertion devices
such as APPLE-type undulators \cite{SASA}, and (ii) X-ray phase retarders \cite{HIRA,FREE}.  The first method is currently being discussed for XFEL projects \cite{tdr-2006}. The second method has not drawn much attention so far, because the retarding plate technique is effective only when the bandwidth of the x-ray beam is sufficiently small not to blur the polarization state. In order to adapt the phase retarder technique to X-ray SASE FEL beamlines it is necessary to add a crystal monochromator to reduce the bandwidth of the incoming beam and to get well-defined circularly-polarized photons. A monochromator can be located in the experimental hall, but the beam intensity available to the phase retarder becomes, in this case, very weak.  Orders of magnitude higher throughput can be achieved when the monochromator is located in the baseline undulator, because the installation of a crystal monochromator in the undulator (for the realization of the self-seeding scheme) always leads to an improvement in terms of spectral flux density.

Our self-seeding setup is very flexible and can be naturally taken
advantage of in different schemes. In particular, it can be used as an ultrafast X-ray pulse measurement setup. The measurement of X-ray pulses on the femtosecond time scale constitutes today an unresolved problem. It is possible to create a few femtosecond X-ray pulses from LCLS in the low-charge mode of operation, but not to measure them.
In \cite{OUR05} we proposed a new method for the measurement of the duration of femtosecond X-ray pulses. In this paper we demonstarte that the self-seeding setup may be used  at LCLS for this purpose too. We also show the advantage of our self-seeding setup when dealing with pump-probe techniques. One of the main technical problem for building femtosecond time-resolved capabilities
is the synchronization between optical pump and hard-X-ray probe pulses with femtosecond accuracy. In this paper we propose a new scheme for optical pump$/$hard X-ray probe femtosecond resolution experiments at the LCLS baseline. The principle of operation of this scheme is essentially based on the exploitation of the time
jitter between optical and hard X-ray pulses. We shift the attention from the problem  of synchronization between pump and probe pulses to the problem of synchronization between two hard X-ray probe pulses with tunable delay and decoupled wavelength. It is proposed to derive both probe hard X-ray pulses  from the same electron bunch but  from different  parts of the LCLS baseline undulator, which has a few percent adjustable gap. This eliminates the need for synchronization and any jitter problem. The delay between the two probe hard X-ray pulses can be introduced in the baseline undulator without the need for dispersive X-ray optical elements (crystals). The idea is to exploit the self-seeding setup, and in particular, for this purpose, the magnetic chicane.

Finally, we note that the LCLS  Far Experimental Hall is an underground structure which containts only three experimental stations. A natural step for expanding the LCLS capabilities would be the addition of a further experimental hall to be built on the surface.  Hard X-ray beams could be distributed to this experimental hall by the LCLS baseline undulator. The main problem in accomplishing this task is that it would require to deflect the full radiation pulse of an angle of order of a radian without perturbation. A high monochromatization of the output radiation constitutes the key for reaching such result. Crystal deflectors located in the LCLS X-ray transport systems would in fact allow the hard X-ray FEL beam to be redirected from the far Experimental Hall to any other hall.

\section{\label{sec:2} Possible self-seeding scheme with a wake monochromator for the LCLS baseline undulator}

\begin{figure}[tb]
\includegraphics[width=1.0\textwidth]{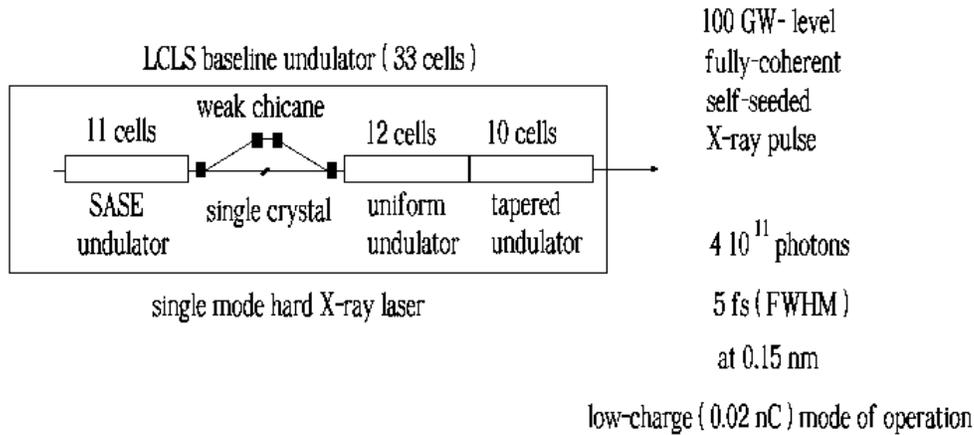}
\caption{Design of the LCLS baseline undulator system for generating of highly monochromatic, high power X-ray pulses. The method exploits a combination of a self-seeding scheme with a single crystal monochromator and an undulator tapering technique. The self-seeding setup is composed of two components, a crystal in Bragg transmission geometry and a 4-m long, weak magnetic chicane. The magnetic chicane accomplishes three tasks by itself. It creates an offset for crystal installation, it removes the electron microbunching produced
in the upstream undulator, and it acts as a delay line for temporal windowing.} \label{lcls1}
\end{figure}

\begin{figure}[tb]
\includegraphics[width=1.0\textwidth]{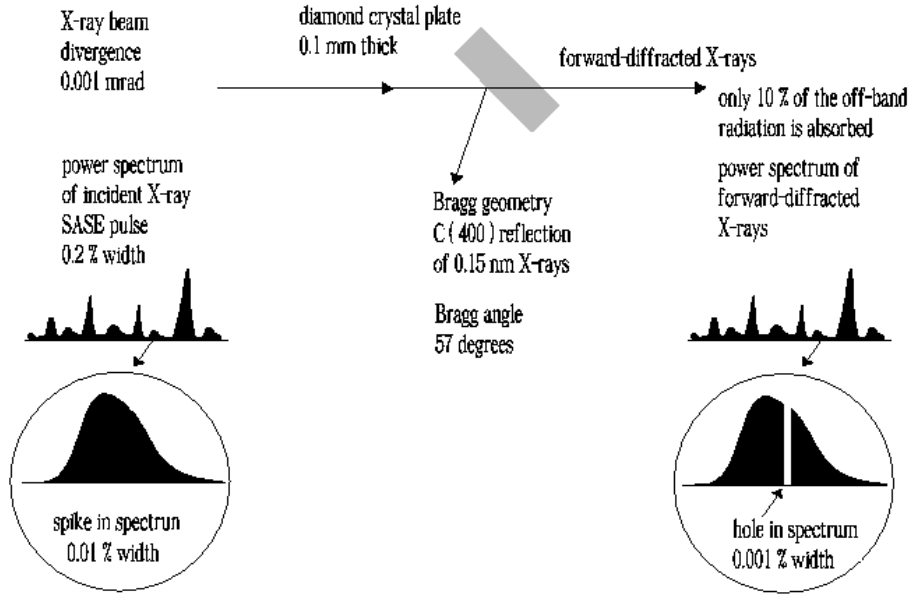}
\caption{Single crystal in Bragg geometry as a bandstop filter for the transmitted X-ray SASE radiation pulse.  The transmittance of the crystal shows a narrow-band absorption resonance when the incident X-ray beam satisfies the Bragg diffraction condition. The temporal waveform of the transmitted radiation pulse is characterized by a long monochromatic wake. After the crystal, the monochromatic wake of the radiation pulse is combined with the delayed electron bunch, and amplified in the downstream undulator.} \label{lcls3}
\end{figure}
\begin{figure}[tb]
\includegraphics[width=1.0\textwidth]{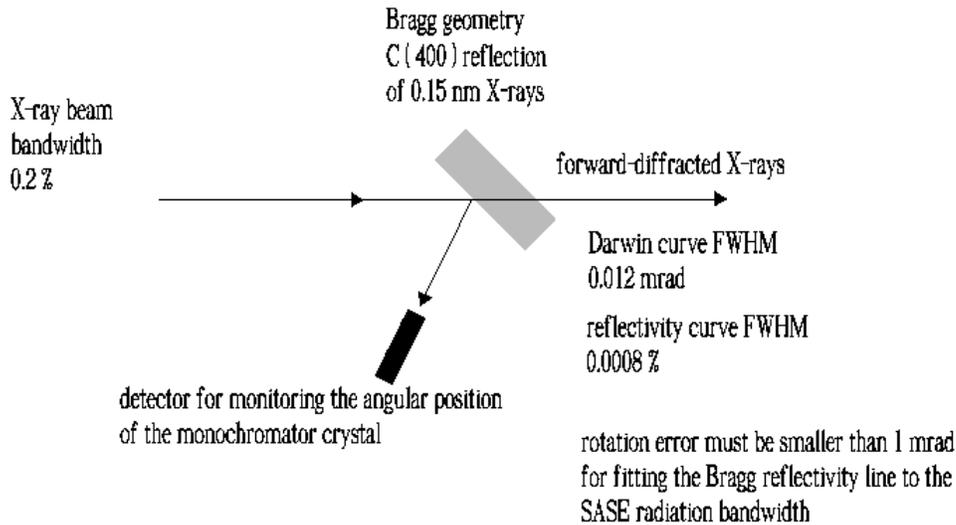}
\caption{Alignment of the wake monochromator crystal.}\label{lcls4}
\end{figure}

\begin{figure}[tb]
\includegraphics[width=1.0\textwidth]{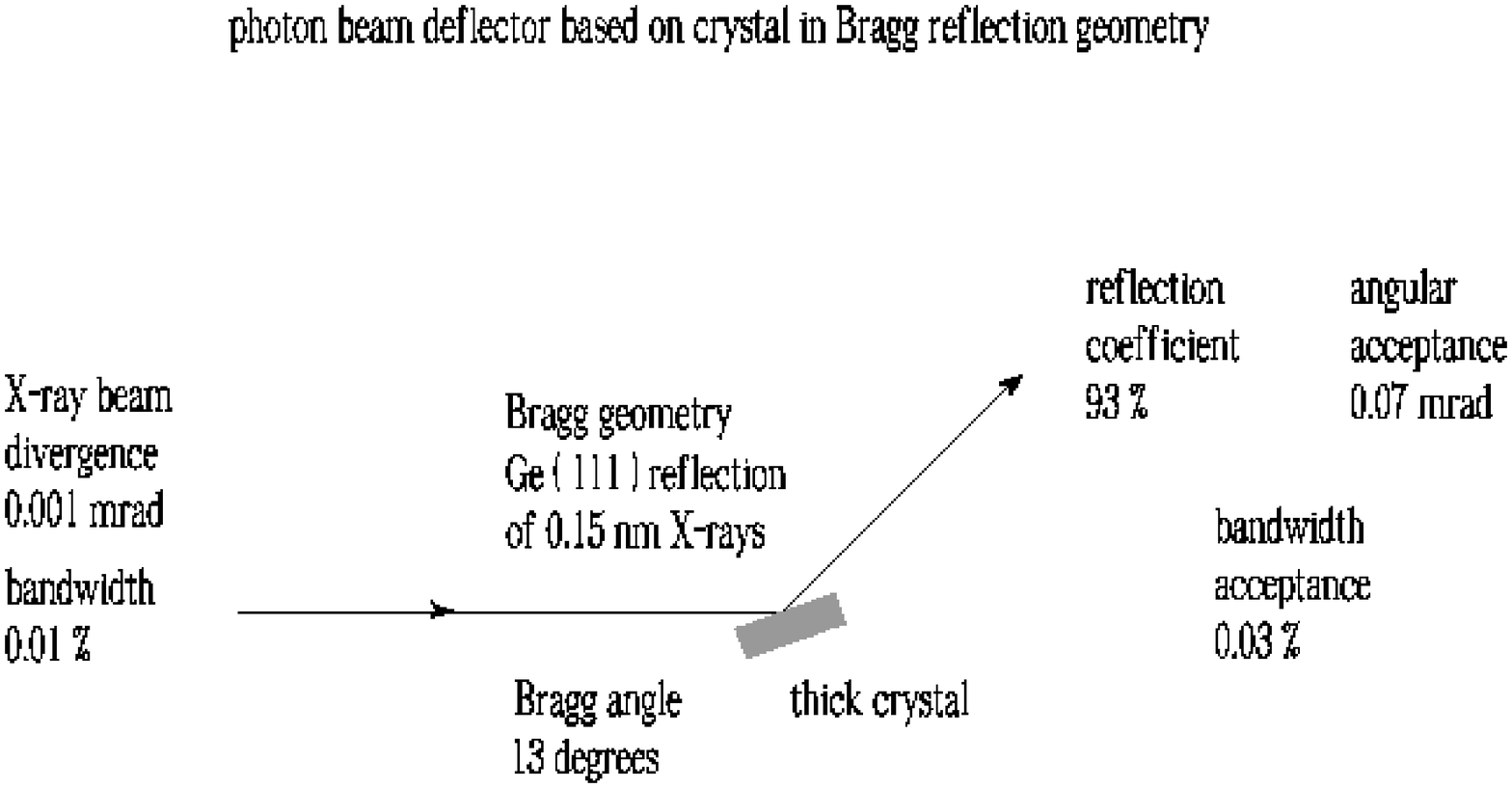}
\caption{Concept of the photon beam deflector for the LCLS X-ray transport system based on the use of a crystal in Bragg reflection geometry.} \label{lcls2}
\end{figure}

In its simplest configuration, a self-seeded FEL consists of an input and an output undulator separated by a monochromator. With reference to Fig. \ref{lcls1} we study the simplest two-undulator
configuration case for a self-seeding setup. The first undulator in Fig. \ref{lcls1} operates in the linear high-gain regime starting from the shot-noise in the electron beam. After the first undulator, the output SASE radiation passes through the monochromator, which reduces the bandwidth to the desired value. A distinguishing feature of our method is that it uses a single crystal in the Bragg transmission geometry, instead of a fixed exit four-crystal monochromator. As a result, the X-ray optics  is extremely simple to align: in involves no beam recombining and no scanning of the delay. According to the wake monochromator principle, the  SASE pulse coming from the first undulator impinges on the crystal set for Bragg diffraction. Then, the single crystal in Bragg geometry operates as a bandstop filter for the transmitted X-ray SASE radiation pulse, as shown in Fig. \ref{lcls3}.  When the incident angle and the spectral contents of  the incoming beam satisfy the  Bragg diffraction  conditions, the temporal waveform of the transmitted radiation pulse shows a long monochromatic wake. The duration of this wake is inversely proportional to the bandwidth of the absorption line in the transmittance spectrum. The alignment tolerance of the crystal angle is expected to be in the range of the mrad, in order to fit the Bragg reflectivity line to the SASE radiation bandwidth, Fig. \ref{lcls4}.

While the radiation is sent through the crystal, the electron beam
passes through a magnetic chicane, which accomplishes three tasks: it creates an offset for the crystal installation, it removes the
electron microbunching produced in the first undulator, and it acts as a delay line for the implementation of a temporal windowing process. In this process, the magnetic chicane shifts the electron bunch on top of the monochromatic wake created by the bandstop filter thus temporally selecting a part of the wake. By this, the electron bunch is seeded with a radiation pulse characterized by a bandwidth much narrower than the natural FEL bandwidth. For the hard X-ray wavelength range, a small dispersive strength $R_{56}$ in the order of a few microns is sufficient to remove the microbunching generated in the first undulator part.  As a result, the choice of the strength of the magnetic chicane only depends on the delay that one wants to introduce between electron bunch  and radiation. The optimal value amounts to $R_{56} \simeq 6 ~\mu$m for the low-charge mode of operation. Such dispersion strength is small enough to be generated by a short ($4$ m-long) magnetic chicane to be installed in place of a single undulator module. Such chicane is, at the same time, strong
enough to create a sufficiently large transverse offset for installing the crystal.

The setup proposed will produce fully coherent X-ray pulses. In order to increase the output power, tapering tachniques are foreseen \cite{TAP1}-\cite{TAP4}. Tapering consists in a slow reduction of the field strength of the undulator in order to preserve the resonance wavelength as the kinetic energy of the electrons decreases due to the FEL process. The undulator taper could be  simply implemented in steps from one undulator module to the next, and is provided by changing the undulator gap. Since here we start the FEL process from a monochromatic seed, rather than from noise, tapering becomes particularly efficient \cite{FAWL,CORN,WANG}.

Fig. \ref{lcls1}  shows the design principle of our setup for high power mode of operation. After the $11$ cell-long SASE input undulator and the monochromator, an output undulator follows, consisting of two sections. The first section is composed by $12$ uniform cells, while tapering is implemented in the second section, consisting of $10$ cells. The monochromatic seed is exponentially amplified by passing through the first uniform section of the output undulator. This section is long enough to reach saturation, and yields an output power of about $15$ GW. Such monochromatic FEL output is finally enhanced up to the $100$ GW-level in the second output-undulator section, by tapering the undulator parameter of about $1 \%$ over the last $10$ cells after saturation.  In summary, our scheme aims to generating  $100$ GW-level, fully-coherent X-ray pulses in the tapered LCLS baseline undulator by means of a self-seeding technique, which produces a highly monochromatic seed.

With the radiation beam monochromatized down to the Fourier transform limit, a variety of different techniques leading to enhance the LCLS
capabilities become feasible. In particular, it is possible to use crystals in Bragg reflection geometry as movable deflectors, Fig. \ref{lcls2}.  For example, for the Ge(111) reflection, the angular acceptance of the crystal deflector is of the order of $70 ~\mu$rad and the spectral bandwidth is about $0.03 \%$. Thus, the angular and frequency acceptance of the deflector are much wider compared to the photon beam divergence and bandwidth, which are respectively of the order of a microradian, and $0.01 \%$ for the low-charge mode of operation. It is therefore possible to deflect the full photon beam without perturbations. About $93 \%$ of the
reflectivity can be achieved for wavelengths around $0.15$ nm. Once
experience with self-seeding setup operations will be available, it will be possible to implement such crystal deflectors for the benefit of the LCLS X-ray transport system.

\section{\label{sec:3} Self-seeding - key to obtain full polarization control at LCLS hard X-ray FEL}

A particularly useful feature becoming easily available with the production of highly monochromatized radiation using self-seeding technique is the full control over the polarization state of the FEL radiation pulse. As mentioned in the Introduction, the idea that we develop in this paper is based on the straightforward use of X-ray phase retarders \cite{HIRA,FREE}, rather than on the installation of special insertion devices. X-ray retarders are birefringent crystals having different indexes of refraction along two orthogonal axis. When the thickness of these crystals is such that the induced phase shift between two orthogonal electric fields pointing along these axis is just $\pi/2$, they behave as quarter wave plates. As is well-known, if a quarter wave plates is adjusted in a way that the axis directions make an angle of $\pi/4$ rad with respect to the linear polarization direction of the incident beam, then the particular phase retardation transforms the linearly polarized field into a circularly polarized field, right-handed or left-handed depending on the sign of the $\pi/2$ phase shift. Birefringence for X-rays occurs in perfect crystals near the Bragg diffraction. The most convenient way to take advantage of such effect is to use the forward-diffracted (i.e. transmitted) beam outside the reflection range.

The ideal performance of X-ray phase retarders is achieved when the
incoming beam has no angular divergence nor photon energy spread. Such an incoming beam has a definite deviation from the exact Bragg condition, hence suffers a definite phase retardation between $\sigma$- and $\pi$-polarization components. The incoming beam from an X-ray SASE FEL is linear polarized and has a sufficiently small angular divergence, but a large photon energy spread, which would blur the polarization state. Thus, in order to adapt the phase retarder to the X-ray SASE FEL beamlines it is necessary to add a crystal monochromator to reduce the bandwidth of the incoming beam and to get a well-defined circular-polarization state. If the monochromator is located in the experimental hall, the beam intensity on the retarder becomes rather weak. It is therefore desirable to install the monochromator already within the baseline undulator. Of course, this situation is automatically realized when our self-seeding scheme is implemented. The self-seeding thus proves to be the key for obtaining full polarization control of the FEL pulse. Quarter wave plates can probably be used to transform the linear polarization delivered by a planar XFEL undulator into circular polarization without any major difficulty. Compared to specially designed undulators delivering circular polarization, there are some advantages in using quarter wave plates: the switching of the polarization handedness is easily done by small rotations of the crystal and therefore very easy to handle, and the cost is much lower than for the corresponding undulator setup. However, quarter wave plates have the heavy disadvantage that they are well-suited for the hard X-ray wavelength range only.

\begin{figure}[tb]
\includegraphics[width=1.0\textwidth]{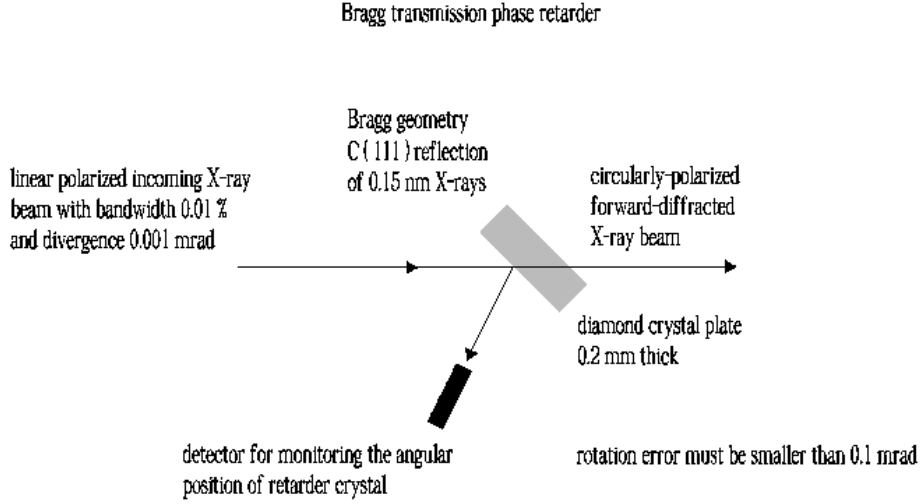}
\caption{Bragg transmission X-ray phase retarder based on a $0.2$ mm thick diamond crystal at 8 keV.  The Diamond (111) reflection is used. A linearly polarized, monochromatic X-ray beam is directly converted to circularly-polarized X-ray beam.} \label{lcls6}
\end{figure}
\begin{figure}[tb]
\includegraphics[width=1.0\textwidth]{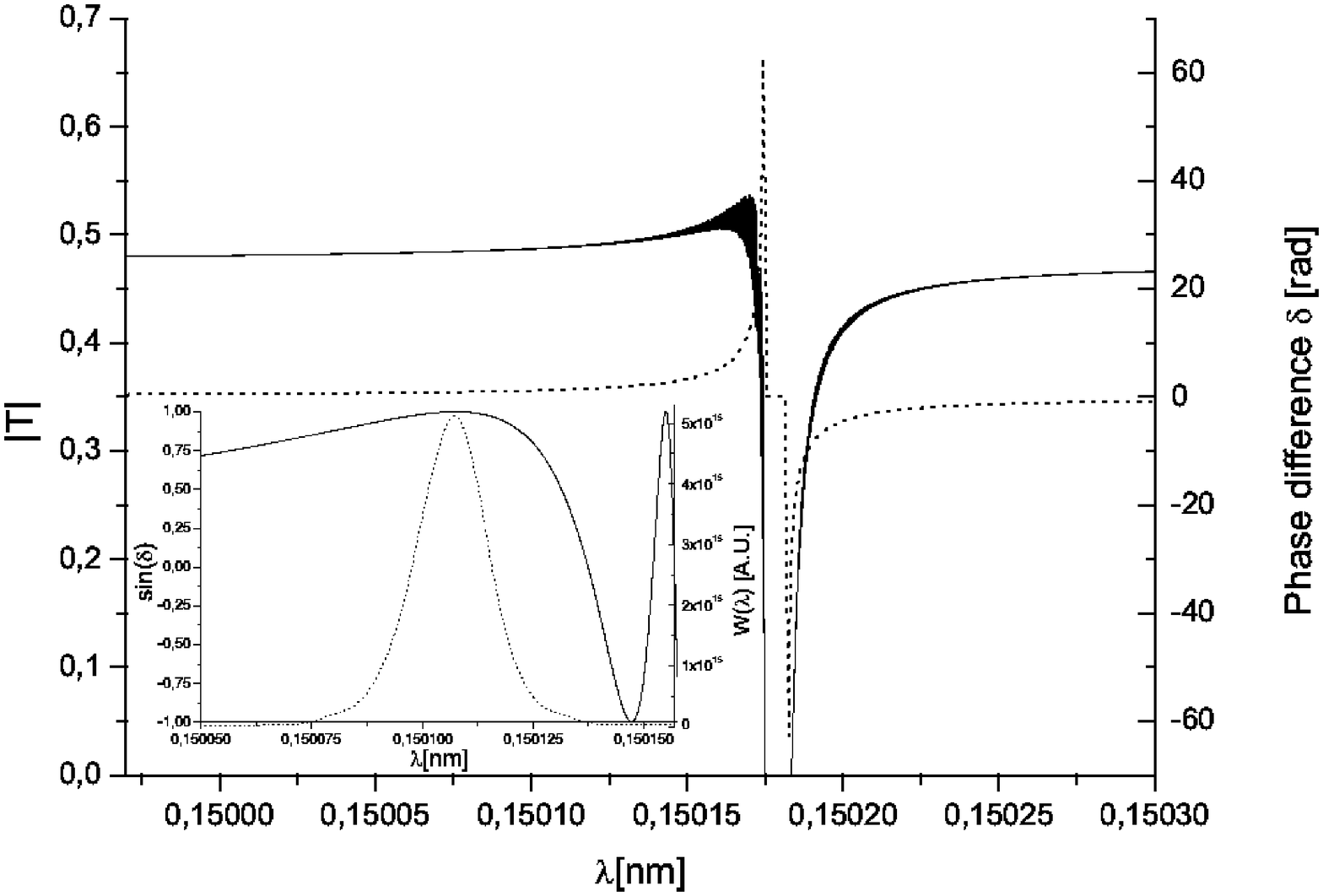}
\caption{Main plot: modulus of the transmittance for the $\sigma$-polarization component relevant to the Bragg (111) reflection of a $0.2$ mm thick diamond crystal at 8 keV (solid line), and phase difference $\delta$ between $\sigma$- and $\pi$-polarization components. Inset: $\sin(\delta)$ (solid line) and average spectrum (dotted line) from the last tapered output undulator. The average degree of circular polarization is about $93 \%$. Phase retardation of $90$ degrees is produced at deviation parameter $\eta \simeq -14$,  where the transmissivities of the $\sigma$- and $\pi$-polarization components are about $50 \%$.} \label{deltaphase}
\end{figure}
For a given reflection and wavelength, one has to optimize the beam path in the crystal with respect to the phase shift and to the transmission. It is generally necessary to use a crystal with as little absorption as possible. Practically, thick diamond crystals may be used as quarter wave plates. Thicker phase retarders of lighter materials become a quarter wave plate far on the tails of the Bragg peak, and phase retardation becomes more insensitive to the bandwidth of the incoming beam. For our calculations we used the symmetric 111 reflection of a diamond crystal in Bragg geometry, $0.2$ mm-thick at 8 keV, combined with the self-seeded LCLS X-ray FEL operated in the low-charge mode, Fig. \ref{lcls6}. Results are shown in Fig. \ref{deltaphase}. The main plot shows the modulus of the transmittance relevant to the Bragg (111) diffraction of X-rays at $0.15$ nm from such diamond crystal ($\sigma$-polarization), and the phase difference between $\sigma$- and $\pi$-polarization components, which is indicated with $\delta$. In Section \ref{sec:5} we discuss the approximation that we used for calculating the phase-difference $\delta$, which is plotted in the inset (solid line), and  its accuracy. In order obtain the maximum degree of circular polarization we need to tilt the crystal in such a way that the maximum of the impinging radiation spectrum coincides with the value $\delta = \pi/2$ on the tail of the phase difference $\delta(\omega)$. The average degree of circular polarization from the transmitted beam is then found as

\begin{eqnarray}
\langle P_c\rangle = \int W(\omega) \sin[\delta(\omega)] d \omega \cdot \left[\int W(\omega)  d \omega\right]^{-1}~,
\label{Pc}
\end{eqnarray}
where $W(\omega)$ is the averaged spectrum and integrals are performed over all frequencies. The resultant maximum $\langle P_c\rangle$ for the case under examination has been estimated to be $0.93$.

In closing this Section, it should be noted that the proposed setup for polarization control is characterized by energy-tunability and helicity-switchability. Wide energy tunability is attainable because the phase retardation is controlled through the deviation angle from the exact Bragg condition. Helicity switching is quite easy to obtain because, as mentioned above, the sign of the phase shift is reversed for the opposite angular sides of the exact Bragg condition.

\section{\label{sec:4} Phase shift of forward-diffracted X-rays}

\begin{figure}[tb]
\includegraphics[width=1.0\textwidth]{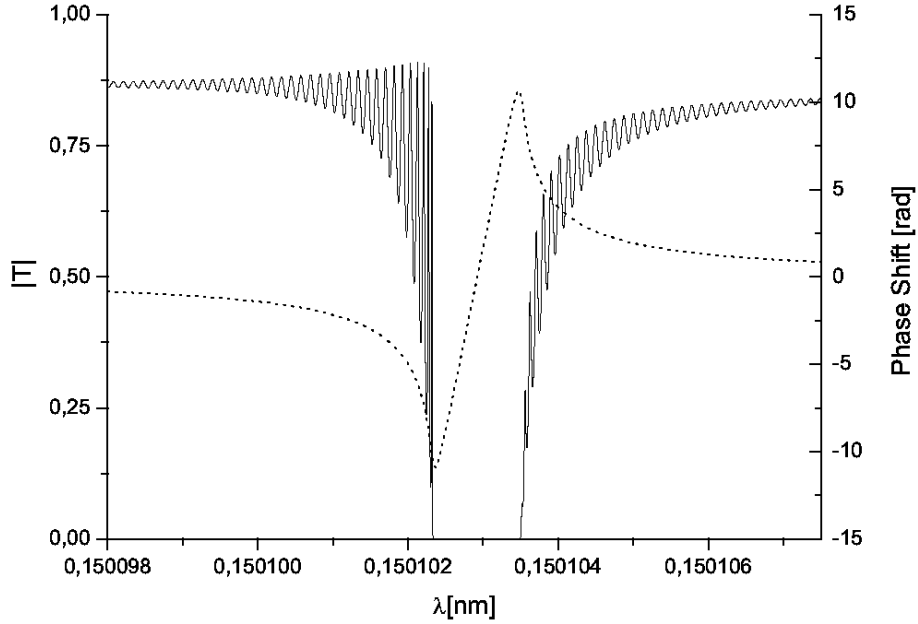}
\caption{Transmissivity (sigma polarization) relevant to the Bragg
(400) diffraction of X-rays at $0.15$ nm from a Diamond crystal with
a thickness of $0.1$ mm. Modulus (solid line) and corresponding phase (dotted line) recovered with the help of the Kramers-Kronig approach.} \label{transm}
\end{figure}

In \cite{OURX, OURY2} we discussed about the relation between modulus and phase of the transmittance of the diamond crystal used in the monochromatization procedure in terms of Kramers-Kronig relations.
The modulus of the transmittance, calculated on the basis of the dynamical theory of diffraction \cite{WECK}, and the phase, recovered with the help of the Kramers-Kronig approach implemented in the software package \cite{LUC2}, are shown in Fig. \ref{transm} for the case of Bragg (400) diffraction of X-rays at $0.15$ nm from a diamond crystal with a thickness of $0.1$ mm. In this Section we present a thorough justification for the relation between modulus and phase.

X-ray multiple scattering in a perfect crystal is described by the dynamical theory of X-ray diffraction (see e.g. \cite{AUTH} for a comprehensive treatment). Consider an X-ray beam incident on a crystal in symmetric Bragg geometry. According to the dynamical theory of diffraction, when the incident angle of radiation is in the vicinity of Bragg's diffraction condition, one can write the modulus of the transmittance in Bragg geometry as in \cite{AUTH}:

\begin{eqnarray}
&&|T(\eta)| = 2 \sqrt{\left|\eta^2-1\right|} \cr && \times \left|
\left(\eta + \sqrt{\eta^2-1}\right)\exp\left[\frac{i \pi t_c}{\Lambda_B} \sqrt{\eta^2-1}\right] - \left(\eta - \sqrt{\eta^2-1}\right) \exp\left[-\frac{i \pi t_c}{\Lambda_B} \sqrt{\eta^2-1}\right]\right|^{-1}\cr &&
\label{modT}
\end{eqnarray}
where $t_c$ is the crystal thickness, $\Lambda_B$ is the extinction length, and $\eta$ is known in literature as the deviation parameter, which best describes the theory results.  When absorption is neglected, $\eta$ is real. Adding a subscript $r$ to stress reality, $\eta$ can be written as

\begin{eqnarray}
\eta_r = (\Delta \theta - \Delta \theta_{0r})/\delta_r~,
\end{eqnarray}
where $2 \delta_r$ is the Darwin width of the reflectivity curve,  $\Delta \theta$ is the angular departure from the Bragg's angle, and $\Delta \theta_{0r}$ is the departure of the middle of the reflection domain from the Bragg's angle. Therefore $(\Delta \theta - \Delta \theta_{0r}) = \eta_r \delta_r$ gives, altogether, the angular departure from the middle of the reflection domain. Usually one is interested in the transmission, or in the reflection curve at a fixed frequency of incident radiation, as a function of the deviation from the Bragg's angle. In this paper, instead, we consider the incident angle of the radiation fixed near the Bragg's angle\footnote{In our case of interest we have an angular divergence of the incident photon beam of about a microradian, which is much smaller than the Darwin width of the rocking curve ($10 - 15 \mu$rad.) As a result, we assume that all frequencies impinge on the crystal at the same angle in the vicinity of the Bragg diffraction condition. Note that mirror reflection takes place for all frequencies and, therefore, the reflected beam has exactly the same divergence as the incoming beam.}, and we scan through the frequency deviation $\Delta \omega$ from the frequency $\omega_B$ associated to the Bragg's angle $\theta_B$ by Bragg's law\footnote{Note that the incidence angle can be tuned once and for all in order to center the rocking curve at the desired frequency by slightly tilting the crystal. This means that we are free to select a given $\omega_B$ within the spectrum of the incoming radiation corresponding to a particular Bragg's angle $\theta_B$.}.  A given fixed incidence angle corresponds to the Bragg's angle for a certain frequency through the Bragg's law. As concerns the value of the deviation parameter, once the incidence angle is fixed deviations in frequency correspond to deviations from the Bragg's angle.  In particular, for $\Delta \theta \ll 1$, Bragg's law allows one to write the relation between $\Delta \theta$ and $\Delta \omega/\omega_B$ as

\begin{eqnarray}
\frac{\Delta \omega}{\omega_B} \longleftrightarrow \Delta \theta \cot(\theta_B)~,
\label{omegeta}
\end{eqnarray}
yielding

\begin{eqnarray}
\frac{\Delta \omega}{\omega_B} = (\delta_r \eta_r + \Delta \theta_{0r}) \cot(\theta_B)~.
\label{omegeta2}
\end{eqnarray}
This means that, aside for a shift given by the term in $\Delta \theta_{0r}$, $\eta_r \propto \Delta\omega$. The previous discussion holds in absence of absorption. When absorption is included, $\eta$, $\Delta \theta_0$ and $\delta$ become complex. In particular $\eta = \eta_r + i \eta_i$ where $\eta_i = A \eta_r + B$, and expressions for $A$ and $B$ (together with those for $\Delta \theta_0$ and $\delta$ and a detailed derivation of results) can be found in \cite{AUTH}.

Having discussed the relation between $\eta$ and $\omega$, one can use Eq. (\ref{modT}) to plot the modulus of the transmissivity as a function of the frequency. At this stage, the phase of the transmissivity remains, however, unknown. A straightforward way of calculating it, is by applying again, explicitly, the dynamical theory of diffraction \cite{AUTH}. In previous works of us \cite{OURX, OURY2}, we chose an alternative derivation, based on the use of Kramers-Kronig relations, which we briefly review here.
Let us regard our crystal as a filter with transmission $T(\omega) = |T(\omega)| \exp[i\Phi(\omega)]$. According to Titchmarsch theorem (see \cite{LUCC} and references therein for a recent review on the subject) causality\footnote{Causality simply
requires that the filter can respond to a physical input after the
time of that input and never before.} and square-integrability of
the inverse Fourier transform of $T(\omega) = |T(\omega)| \exp[i
\Phi(\omega)]$, which will be indicated with $\mathcal{T}(t)$, is equivalent\footnote{Here we are tacitly assuming, as required by Titchmarsh theorem, that $T(\omega)$ is a square integrable function. This is physically reasonable, since there will always be some cutoff frequency for which the filter does not transmit anymore. Strictly speaking, the mathematical model of $T$ in Eq. (\ref{modT}) is not square integrable. One should therefore understand a certain cutoff frequency in the definition in Eq. (\ref{modT}). The presence of such cutoff does not change the following considerations.} to the existence of an analytic continuation of $T(\omega)$ to
$\Omega = \omega + i \omega'$ on the upper complex $\Omega$-plane
(i.e. for $\omega'>0$). The same theorem also shows that the two
previous statements are equivalent to the fact that real and
imaginary part of $T(\omega)$ are connected by Hilbert
transformation. Since $\mathcal{T}(t)$ must be real (thus implying that $T^*(\omega)=T(-\omega)$), from the Hilbert transformation follows the well-known Kramers-Kroninig relations
\cite{KRAM,KRON}:

\begin{eqnarray}
&&\mathrm{Re}[T(\omega)]=\frac{2}{\pi}\mathcal{P}
\int_0^{\infty}
\frac{\omega'\mathrm{Im}[T(\omega')]}{\omega'^2-\omega^2}
d\omega' \cr &&
\mathrm{Im}[T(\omega)]=-\frac{2}{\pi}\mathcal{P}
\int_0^{\infty}
\frac{\mathrm{Re}[T(\omega')]}{\omega'^2-\omega^2}
d\omega'~,
\label{KKrel}
\end{eqnarray}
linking real and imaginary part of $T(\omega)$. A similar
reasoning can be done for the modulus $|T(\omega)|$ and the phase
$\Phi(\omega)$, see \cite{TOLL}. In fact, one can write

\begin{eqnarray}
\mathrm{ln}[T(\omega)] = \mathrm{ln}[|T(\omega)|] + i
\Phi(\omega)~.\label{ln}
\end{eqnarray}
Note that $T^*(\omega)=T(-\omega)$ implies that
$|T(\omega)|=|T(-\omega)|$ and that $\Phi(\omega) = -
\Phi(-\omega)$. Therefore, using Eq. (\ref{ln}) one also has that
$\mathrm{ln}[T(\omega)]^*=\mathrm{ln}[T(-\omega)]$.
Then, similarly as before, application of Titchmarsh theorem shows
that the analyticity of $\mathrm{ln}[|T(\Omega)|]$ on the upper
complex $\Omega$-plane implies that

\begin{eqnarray}
\Phi(\omega)=-\frac{2}{\pi}\mathcal{P} \int_0^{\infty}
\frac{\mathrm{ln}[T(\omega')] }{\omega'^2-\omega^2} d\omega'~,
\label{KKrel2}
\end{eqnarray}
%
A direct use of Eq. (\ref{KKrel2}), with $|T|$
given as in Eq. (\ref{modT}), should yield back the phase
$\Phi(\omega)$. As is well known however, in applying
such procedure one tacitly assumes that $\mathrm{ln}[T(\Omega)]$ is
analytical on the upper complex $\Omega$-plane. While causality
implies this fact for $T(\Omega)$, it does not imply it automatically for $\mathrm{ln}[|T(\Omega)|]$. In fact, such function is
singular where $T(\Omega)=0$. If $T(\Omega)$ has zeros on the upper complex plane, these zeros would contribute adding extra terms to the total phase. For this reason, Eq. (\ref{KKrel2}) is known as \emph{minimal} phase solution.

In the following we show that in our case $T(\Omega)$ does not have zeros on the upper complex plane, and that the minimal phase solution is actually the complete solution to the phase problem. To this end, it is sufficient to study the zeros of $|T|$, where $|T|$ is given in Eq. (\ref{modT}) as a function of the complex $\eta$ parameter. In fact, from Eq. (\ref{omegeta}) and from $\eta = (1 + i A) \eta_r + i B$ follows that

\begin{eqnarray}
\eta(\omega) = (1+ i A) \left[\frac{\omega - \omega_B}{\omega_B \delta_r \cot(\theta_B)} - \frac{\Delta \theta_{0r}}{\delta_r}\right] + i B~.
\label{eta}
\end{eqnarray}
Consider now the analytic continuation of $|T(\eta(\omega))|$ to the upper complex $\Omega$-plane. In passing, note that  $|T(\eta(\Omega))|$ must be analytic on such half plane, because the filter is causal, meaning that $\mathcal{\tau}(t) = 0$ for $t<0$. We have to show that $|T(\eta(\Omega))|$ has no zeros on the complex half plane. In order to demonstrate this, it is sufficient to demonstrate that $|T(\eta)|$ has no zero \emph{in the entire complex $\eta$-plane}. First note that the numerator of Eq. (\ref{modT}) becomes zero for $\eta = \pm 1$ only. However, in these  points, the denominator becomes zero too. Using, e.g. de l'Hopital rule one can show that the limit for $\eta \longrightarrow \pm 1$ is finite. Finally, the denominator never goes to infinity (except for $|\eta| \longrightarrow \infty$). We can therefore conclude that the analytic continuation of $|T(\eta(\omega))|$  has no zeros in the upper $\Omega$-plane. Therefore Titchmarsch theorem applies, and the minimal phase solution considered in previous papers \cite{OURX,OURY2} is the correct one.

\section{\label{sec:5} Theoretical background of phase retarders in Bragg transmission geometry}

In this Section we justify the treatment of phase retarders proposed in Section \ref{sec:3}, with reference to \cite{HIRA} and \cite{FREE}. In fact, results in Section \ref{sec:3} are based on the thick-crystal approximated formula given in \cite{HIRA} to calculate the phase difference between $\sigma$- and $\pi$-polarization components. Here we will introduce the thick-crystal approximation for such phase difference and show that, albeit not accounting to absorption, it constitutes a good approximation for our purposes. We have not found discussions about the accuracy of this approximated formula anywhere in published literature. We will limit ourselves to showing that in a few examples, the simplified approach gives the same results as the exact expression for the phase of the transmittance.

First it should be noted that the only relevant quantity needed in order to calculate the degree of circular polarization is the phase difference between the crystal transmittance in the $\sigma$ and $\pi$ polarization directions. This fact can be easily seen considering the definition of the circular degree of polarization $P_c$ in terms of Stokes parameters $S_0$ and $S_3$:

\begin{eqnarray}
P_c = \frac{S_3}{S_0}~.
\label{cirpol}
\end{eqnarray}
Here

\begin{eqnarray}
&&S_0 = \Phi(\hat{x}) + \Phi(\hat{y}) \cr &&
S_3   = \Phi\left[\frac{1}{\sqrt{2}}\left(\hat{x}+i\hat{y}\right)\right] - \Phi\left[\frac{1}{\sqrt{2}}\left(\hat{x}-i\hat{y}\right)\right] \cr &&
\Phi(\hat{n}) = \left|E_x n_x+ i E_y n_y\right|^2~,
\label{cirpol}
\end{eqnarray}
where $\hat{n}$ is any unitary complex vector, while $E_x$ and $E_y$ are the complex electric field components, in the frequency domain, along the $x$ and the $y$ directions. Introducing notations $I_{x,y}$ for intensities and $\phi_{x,y}$ for phases, so that  $E_{x,y} = \sqrt{I_{x,y}} \exp{i \phi_{x,y}}$, and working out elementary calculations, one finds that

\begin{eqnarray}
P_c = - \frac{2 \sqrt{I_x I_y}}{I_x+I_y} \sin(\delta)~,
\label{cirpol2}
\end{eqnarray}
where we further introduced notation $\delta = \phi_{x}-\phi_{y}$ for the phase difference. $P_c$ spans from $-1$ (fully circularly polarized radiation, right handed helicity) to $+1$ (fully circularly polarized radiation, left handed helicity). In order to obtain $P_c= \pm 1$, one needs to have $I_x = I_y$ and $\sin(\delta) = \mp 1$, meaning $\delta = \mp \pi/2$. If the $\pi$- and $\sigma$-polarization components are characterized by the same absorption, the former requirement can be fulfilled orienting the crystal in such a way that both the $\hat{x}$ and $\hat{y}$ direction form an angle of $\pi/4$ with respect to the direction of polarization of the impinging light, which is linearly polarized. Otherwise, such orientation will differ from $\pi/4$. When $I_x = I_y$, Eq. (\ref{cirpol2}) reduces to $P_c = - \sin(\delta)$.

\begin{figure}[tb]
\includegraphics[width=1.0\textwidth]{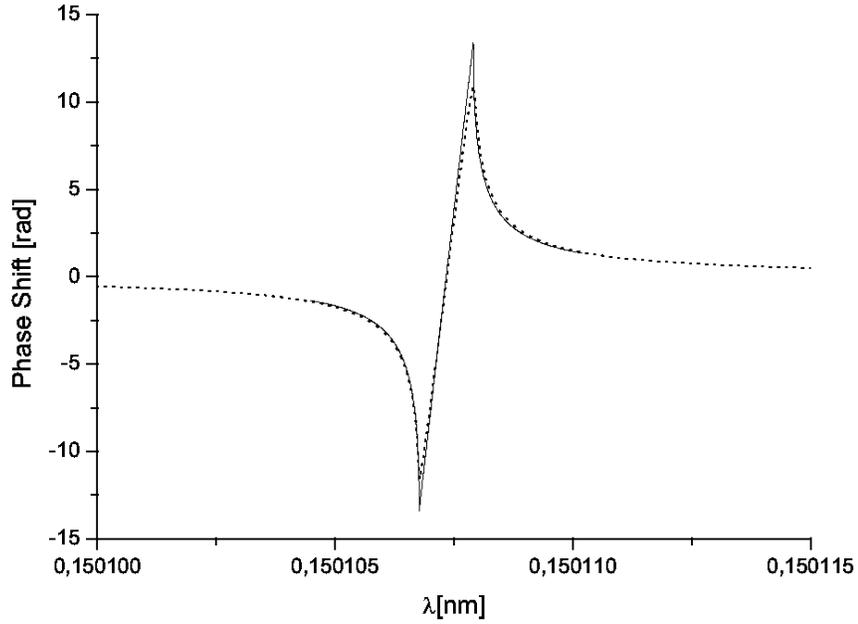}
\caption{Phase relevant to the Bragg (400) diffraction of X-rays at $0.15$ nm from a Diamond crystal with a thickness of $0.1$ mm, $\sigma$-polarization component. The phase retrieved according to the minimal phase method (dotted line) is compared with the phase obtainable with the help of Eq. (\ref{phix})} \label{phases}
\end{figure}

The next step is to find out an expression for $\delta$. In the limit for a large thickness of the crystal, the transmittance is given by \cite{AUTH}:

\begin{eqnarray}
T = \exp\left[\frac{i \pi t_c}{\Lambda_B}\left(\eta - \sqrt{\eta^2-1}\right)\right]~,
\label{Tthick}
\end{eqnarray}
where $t_c$ is the crystal thickness, and $\Lambda_B$ is the extinction length. In this case the phase of the transmittance is just

\begin{eqnarray}
\phi = \frac{\pi t_c}{\Lambda_B} \mathrm{Re}\left[\eta - \mathrm{sign}[\mathrm{Re}(\eta)] \sqrt{\eta^2 -1}\right]~,
\label{phix}
\end{eqnarray}
As an example, a comparison between Eq. (\ref{phix}) and the exact phase relevant to the Bragg 400 diffraction of X-rays at $0.15$ nm from a Diamond crystal with a thickness of $0.1$ mm, already plotted in Fig. \ref{transm}, is given in Fig. \ref{phases} for the $\sigma$-polarization component. We hold the accuracy of Eq. (\ref{phix}) as sufficient for our purposes.

It is important to note that Eq. (\ref{Tthick}) and Eq. (\ref{phix}) apply to both $\sigma-$ and $\pi-$ polarization components. However,  one has \cite{AUTH} that $\Lambda_{B\sigma} = \cos(2 \theta_B) \Lambda_{B\pi}$ and $\delta_{r\sigma} = \delta_{r\pi}/\cos(2\theta_B)$, where subscripts $\sigma$ and $\pi$ indicate the different polarization components. It follows that $\eta$, which depends on $\delta_r$ as $\sim \delta_r^{-1}$, also depends on the polarization component as $\eta_\sigma = \cos(2\theta_B) \eta_\pi$. Therefore,  the linear term in Eq. (\ref{phix}) is the same for both polarization components and cancels when one calculates the phase difference $\delta$. It follows that

\begin{eqnarray}
\delta = \phi_x - \phi_y= \frac{\pi t_c}{\Lambda_{B\sigma}} \mathrm{sign}[\mathrm{Re}(\eta_\sigma)]\mathrm{Re} \left[\sqrt{\eta_\sigma^2 -1}-\sqrt{\eta_\sigma^2 -\cos^2(2 \theta_B)}\right]~,
\label{delta}
\end{eqnarray}
which is the expression used in Section \ref{sec:3} to obtain the results in Fig. \ref{deltaphase}.

\section{\label{sec:6} FEL simulations}

Following the introduction of the proposed method and the previous theoretical Sections, in the present Section we report on a
feasibility study of the single-bunch self-seeding scheme with a
wake monochromator for the LCLS. This feasibility study is performed with the help of the FEL code GENESIS 1.3 \cite{GENE}, running on a parallel machine, for the low-charge mode of operation ($0.02$ nC) corresponding to a bunch length of about $6$ fs. Parameters are presented in Table \ref{tt1}.

\begin{table}
\caption{Parameters for the low-charge mode of operation at LCLS used in
this paper.}

\begin{small}\begin{tabular}{ l c c}
\hline & ~ Units &  ~ \\ \hline
Undulator period      & mm                  & 30     \\
K parameter (rms)     & -                   & 2.466  \\
Wavelength            & nm                  & 0.15   \\
Energy                & GeV                 & 13.6   \\
Charge                & nC                  & 0.02 \\
Bunch length (rms)    & $\mu$m              & 1    \\
Normalized emittance  & mm~mrad             & 0.4    \\
Energy spread         & MeV                 & 1.5   \\
\hline
\end{tabular}\end{small}
\label{tt1}
\end{table}

\subsection{SASE undulator and wake monochromator}

\begin{figure}[tb]
\includegraphics[width=1.0\textwidth]{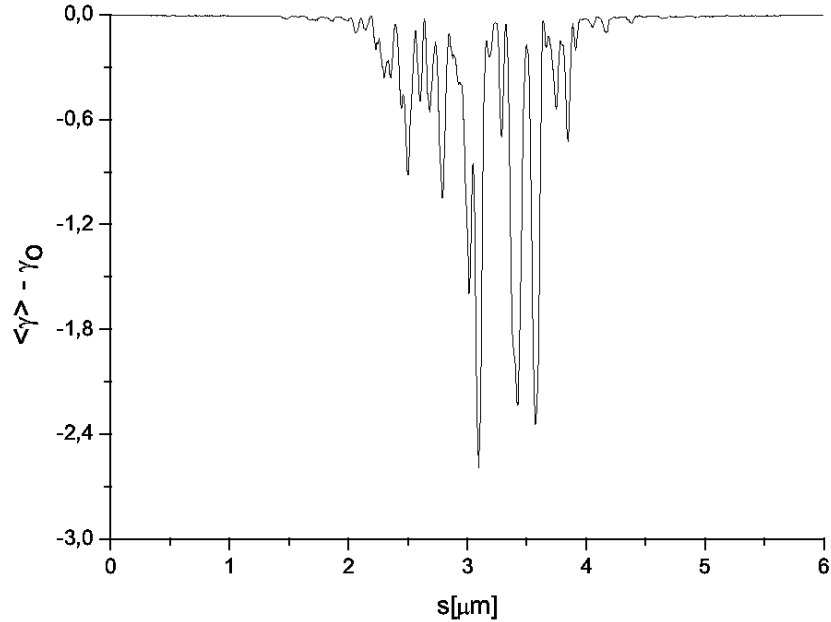}
\caption{Electron beam energy loss after the SASE undulator (11 cells). The data refers to a single typical realization.} \label{Enloss1}
\end{figure}

\begin{figure}[tb]
\includegraphics[width=1.0\textwidth]{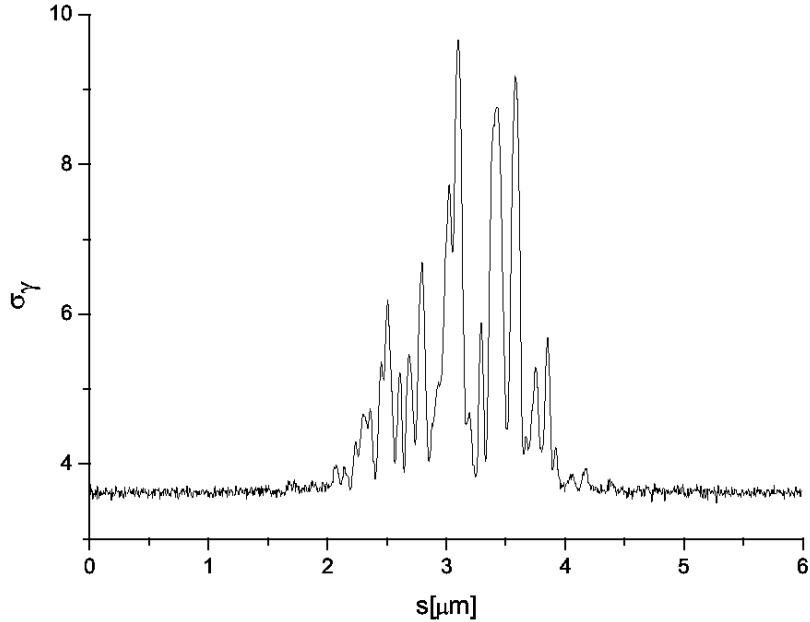}
\caption{Induced energy spread after the SASE undulator (11 cells). The data refers to a single typical realization.} \label{Enspr1}
\end{figure}

\begin{figure}[tb]
\includegraphics[width=1.0\textwidth]{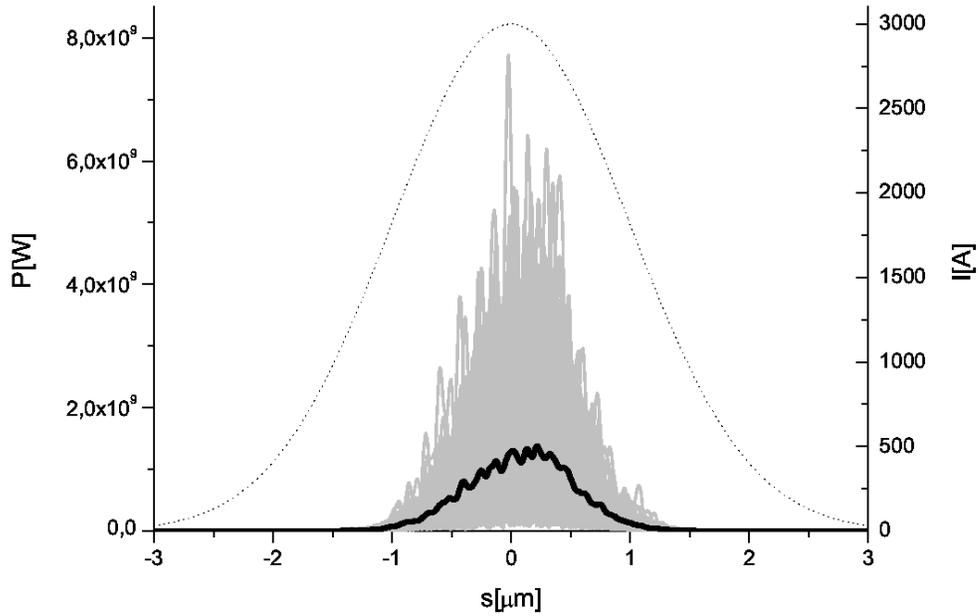}
\caption{Power distribution after the SASE undulator (11 cells). Grey lines refer to single shot realizations, the black line refers to the average over a hundred realizations. The dotted line represents the electron bunch current profile.} \label{1Pin}
\end{figure}
%


\begin{figure}[tb]
\includegraphics[width=1.0\textwidth]{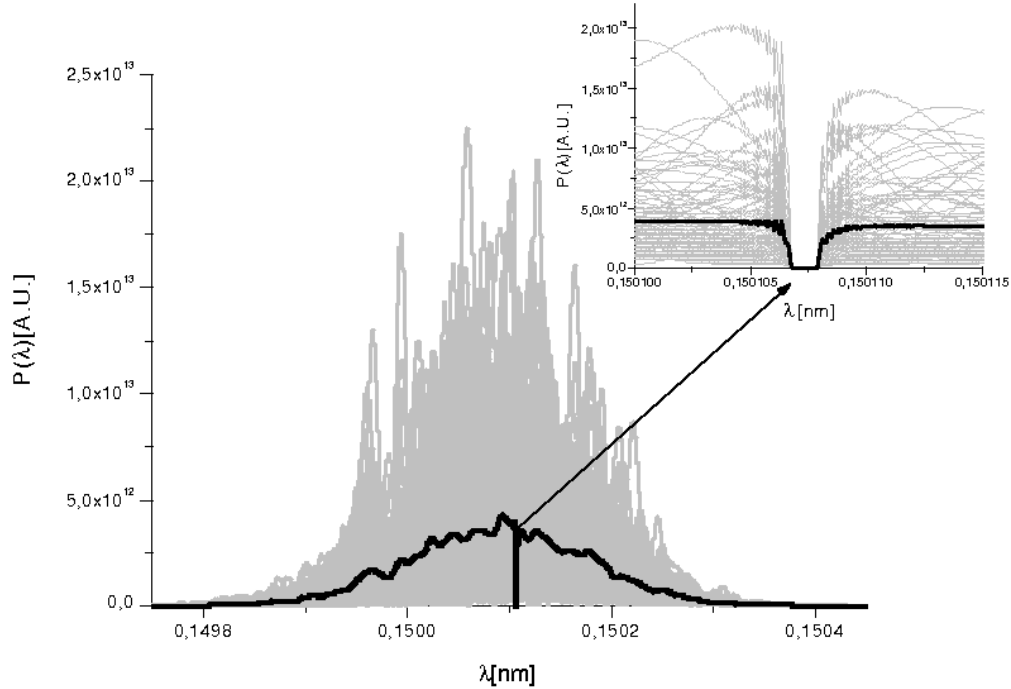}
\caption{Spectrum after the diamond crystal. The bandstop effect is clearly visible. Grey lines refer to single shot realizations, the black line refers to the average over a hundred realizations.} \label{1Spout}
\end{figure}

\begin{figure}[tb]
\includegraphics[width=1.0\textwidth]{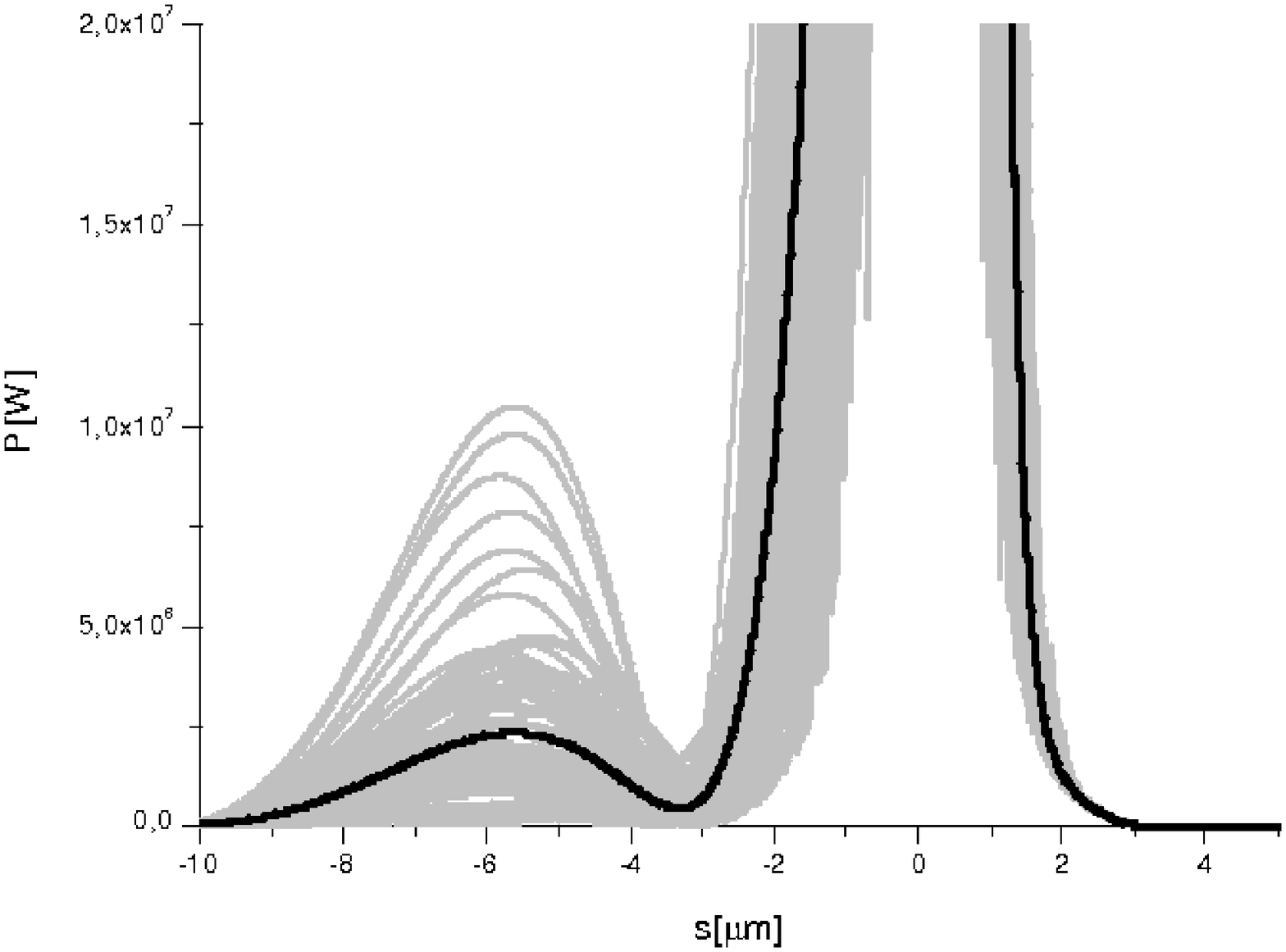}
\caption{Power distribution after the diamond crystal. The
monochromatic tail due to the transmission through the bandstop
filter is now evident on the left of the figure. Grey lines refer to single shot realizations, the black line refers to the average over a hundred realizations.} \label{1Pout}
\end{figure}

With reference to Fig. \ref{lcls1}, the electron beam first passes the 11 cells of the SASE undulator. The FEL is still in the linear regime, so that energy losses and energy spread induced by the FEL process are small enough. As a result, the electron beam can still lase, see Fig. \ref{Enloss1} and Fig. \ref{Enspr1}. The output power is shown in Fig. \ref{1Pin}. 

Following the SASE undulator, the photon pulse is filtered through the wake monochromator, while the electron beam is sent through the short chicane. As explained before, the crystal acts as a bandstop filter. Such effect is best seen in terms of the spectrum in Fig. \ref{1Spout}, where we show a comparison between spectra before
and after the filter. The effect is highlighted in the inset. The
corresponding power is shown in Fig. \ref{1Pout}. As discussed
before, monochromatization does not take place in the frequency
domain. At first glance, the passage through the bandstop filter
is only responsible for slight changes in the power distribution
along the pulse. However, a zoom of the vertical axis shows what
we are interested in: a long, monochromatic tail in the power
distribution on the left side of the picture, Fig. \ref{1Pout}.
Note that there is no corresponding tail on the right side of Fig.
\ref{1Pout}. As already noticed in \cite{OURX}, this fact is consistent with causality.

\subsection{\label{subfirst} First output undulator}

\begin{figure}[tb]
\includegraphics[width=1.0\textwidth]{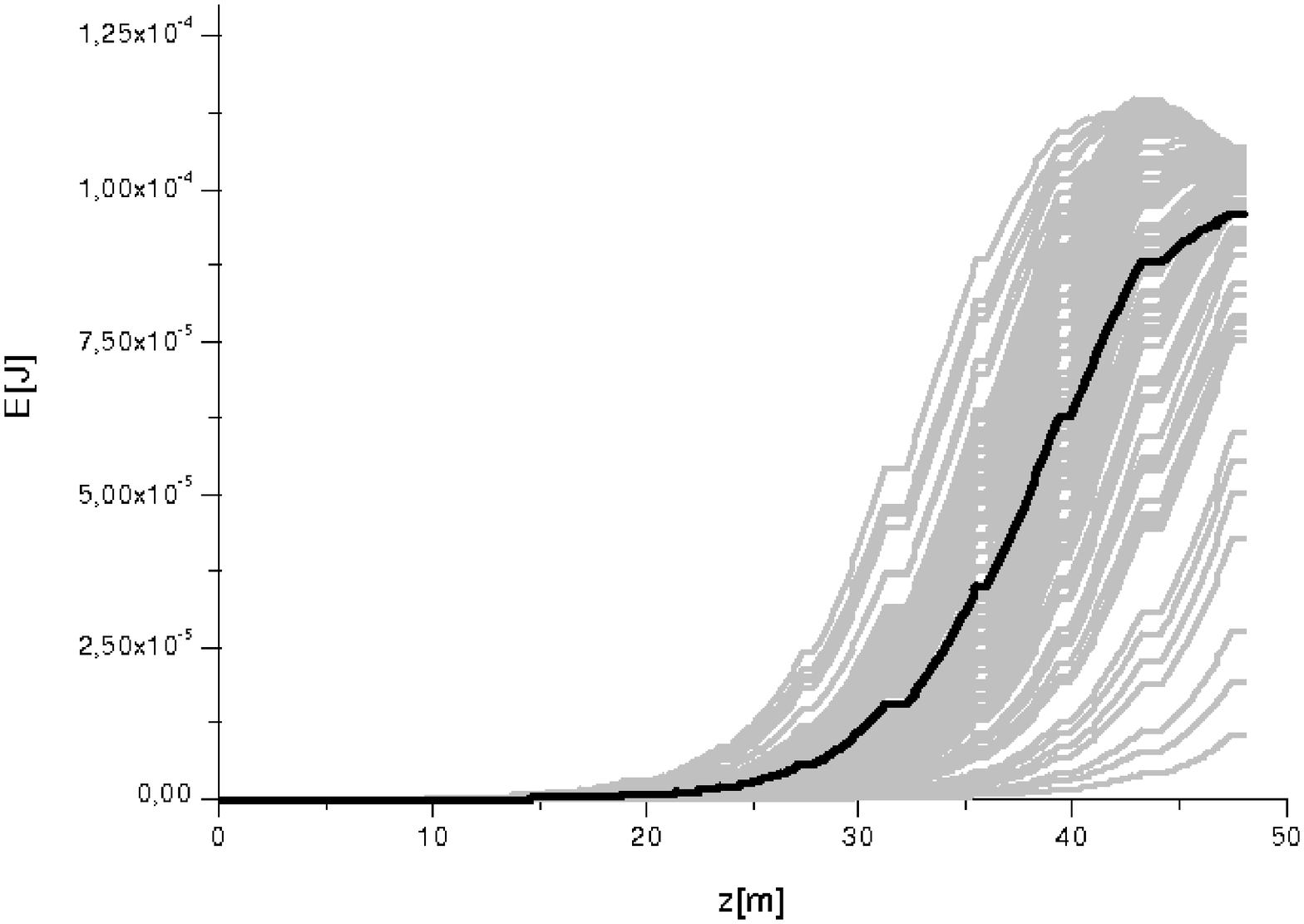}
\caption{Energy in the X-ray radiation pulse versus the length of the uniform output undulator. Here the output undulator
is 12-cells long. Grey lines refer to single shot realizations, the black line refers to the average over a hundred realizations.} \label{2Pz}
\end{figure}

\begin{figure}[tb]
\includegraphics[width=1.0\textwidth]{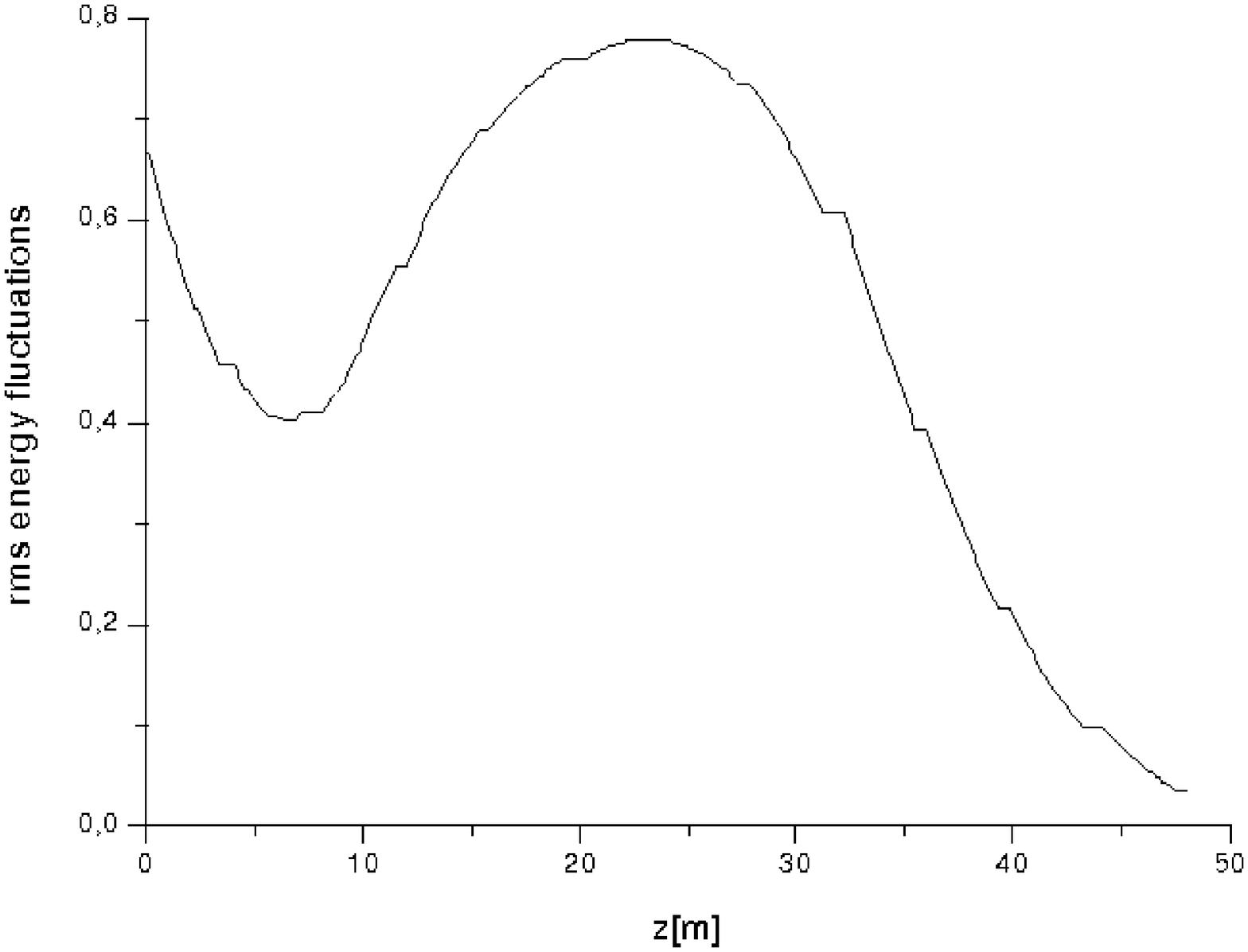}
\caption{rms energy deviation from the average as a function of the
distance inside the uniform output undulator. Here the output undulator is 12 cells-long.} \label{2rms}
\end{figure}
Photons and electrons are recombined after the crystal, so that the electron bunch temporally selects a part of the long wake in Fig. \ref{1Pout}, and is seeded by it. The seeded electron beam goes through the first output undulator, Fig. \ref{lcls1}, which is sufficiently long (12 cells) to reach saturation. Fig. \ref{2Pz} shows the output energy as a function of the position inside the undulator, while in Fig. \ref{2rms} we plot the rms energy fluctuations, as a function of the position inside the undulator as well. As noted e.g. in \cite{OURY2}, at the beginning of the undulator the fluctuations of energy per pulse apparently drop. This can be explained considering the fact that the Genesis output consists of the total power integrated over the full grid up to an artificial boundary, i.e. there is no spectral selection. Therefore, our calculations include a relatively large spontaneous emission background, which has a much larger spectral width with respect to the amplification bandwidth and which fluctuates negligibly from shot to shot. Since there is a long lethargy of the seeded radiation at the beginning of the FEL amplifier, one observes an apparent decrease of fluctuations. Then, when lethargy ends, the seed pulse gets amplified and fluctuations effectively return to about the same level as after monochromator.

\begin{figure}[tb]
\includegraphics[width=1.0\textwidth]{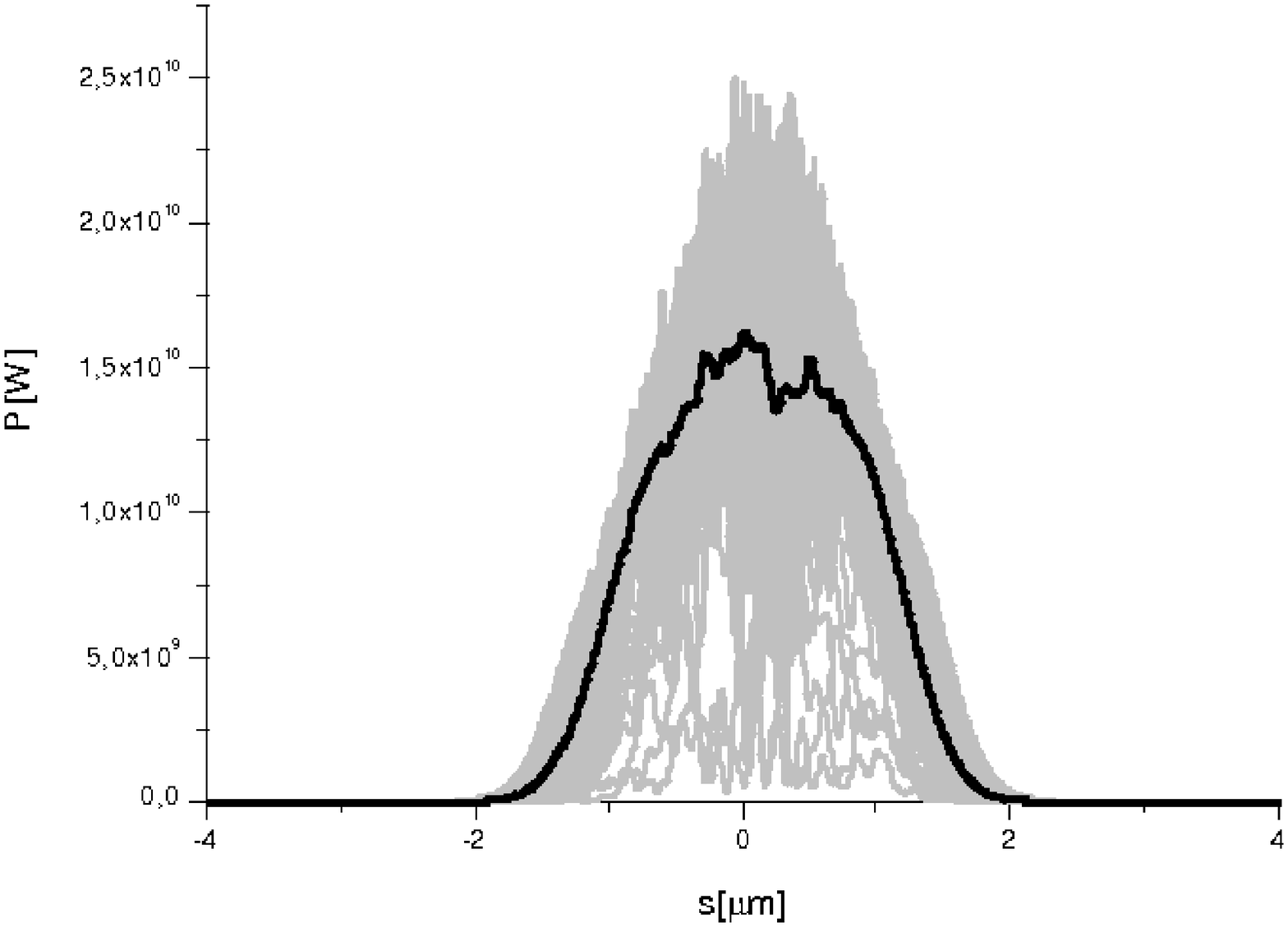}
\caption{Power distribution of the X-ray radiation pulse at saturation without tapering. Here the output undulator is 12-cells long. Grey lines refer to single shot realizations, the black line refers to the average over a hundred realizations.} \label{2Pout}
\end{figure}

\begin{figure}[tb]
\includegraphics[width=1.0\textwidth]{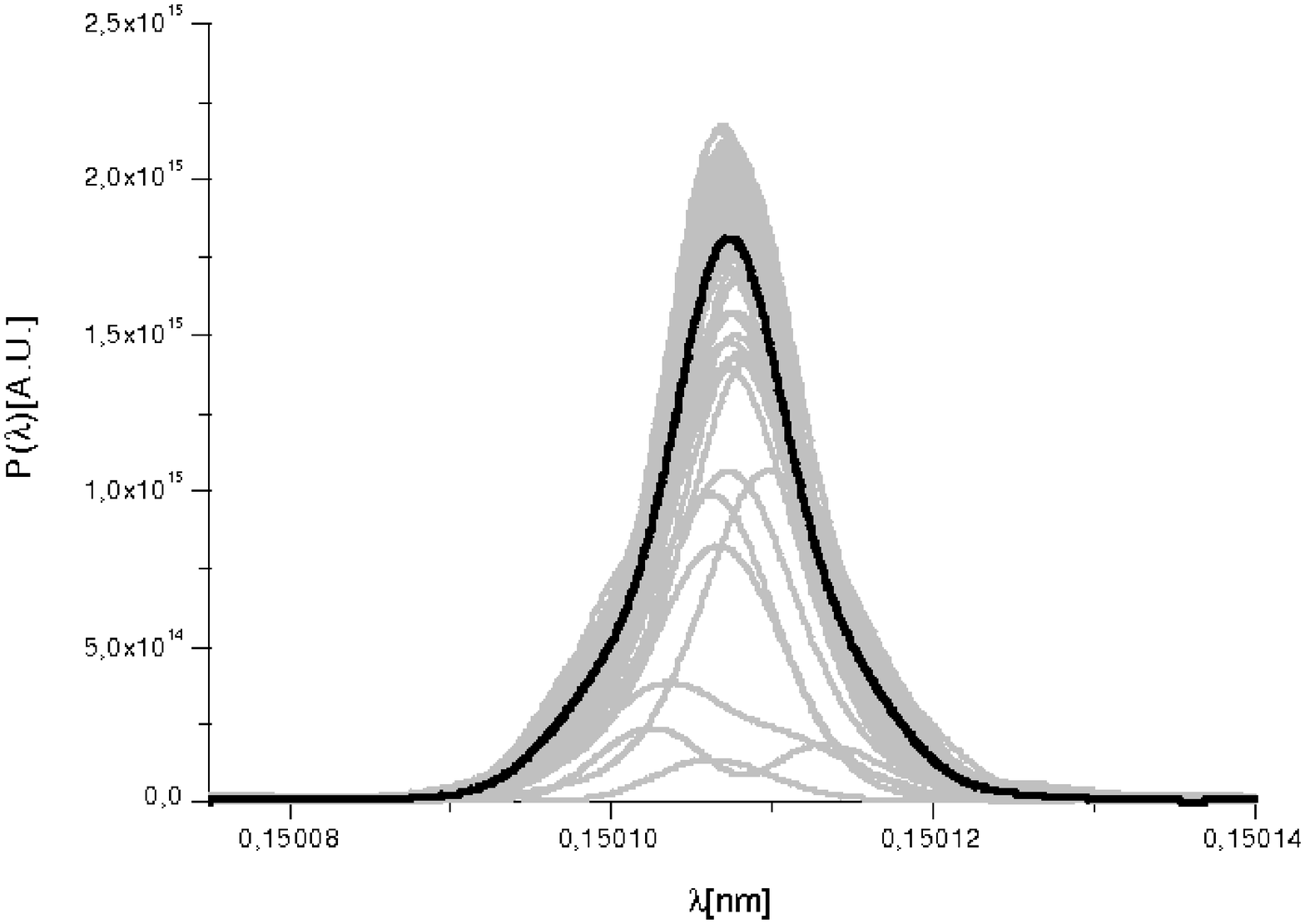}
\caption{Spectrum  of the X-ray radiation pulse at saturation without tapering. Here the output undulator is 12-cells long. Grey lines refer to single shot realizations, the black line refers to the average over a hundred realizations.} \label{2Spout}
\end{figure}
The output power and spectra are shown in Fig. \ref{2Pout} and Fig. \ref{2Spout} respectively. A peak power of about $15$ GW is foreseen. Monochromatization is in the order of $10^{-4}$.

\subsection{Tapering section}

\begin{figure}[tb]
\includegraphics[width=1.0\textwidth]{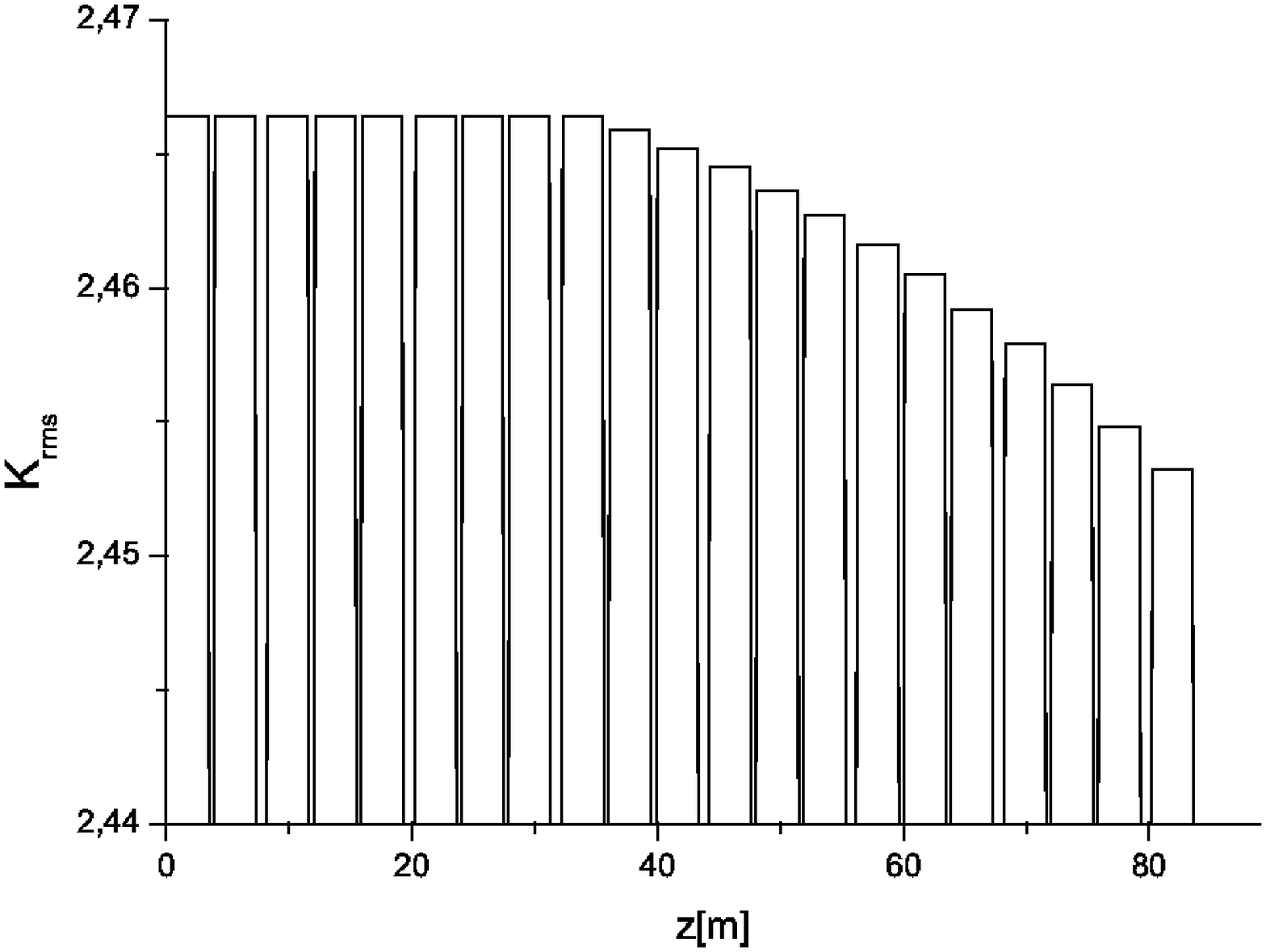}
\caption{Taper configuration for high-power mode of operation at $0.15$ nm} \label{Taplaw}
\end{figure}
Finally, the tapering section follows. The simulations from Section \ref{subfirst} where made by choosing  the number of uniform undulator sections in order to optimize the output characteristics. Similarly, in the simulations including tapering, both an optimized number of uniform and tapered sections have been chosen. The total number of section from the first input undulator to the output is $32$, and we used 9 uniform cells and 12 tapered cells.

The value of the rms $K$ parameter in Table \ref{tt1} refers to the untapered sections. The optimal\footnote{Note that additional tapering should be considered to keep undulators tuned in the presence of
energy loss from spontaneous synchrotron radiation. In Fig. \ref{Taplaw} we present a tapering configuration which is to be considered as an addition to this energy-loss compensation
tapering.} tapering configuration is presented, instead, in Fig. \ref{Taplaw}.  Such law has been determined phenomenologically, with the help of numerical experiments. The effective value of K is kept constant through each  undulator segment. In other words, tapering consists of a stepwise change of K from segment to segment (segment
tapering).

\begin{figure}[tb]
\includegraphics[width=1.0\textwidth]{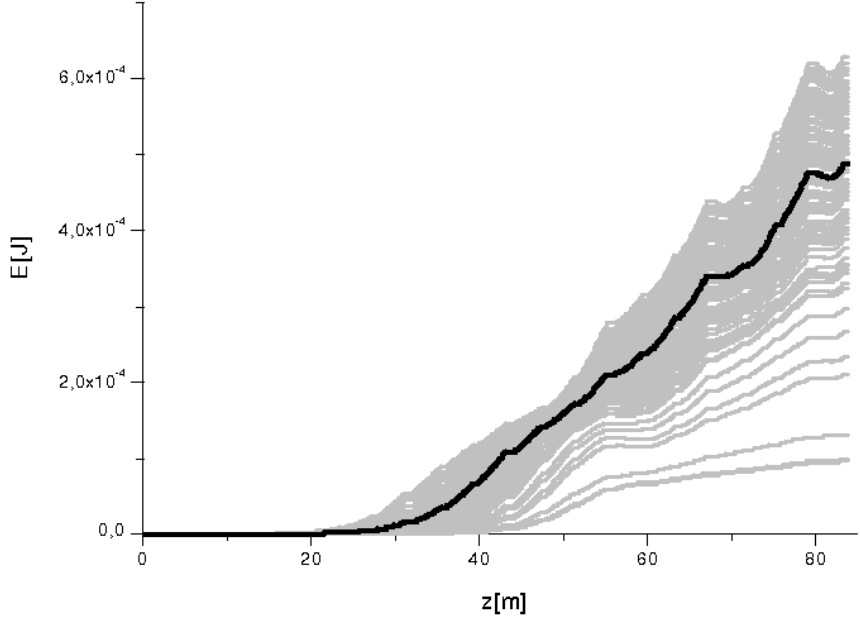}
\caption{Energy in the X-ray radiation pulse versus the length of
the output undulator in the case of undulator tapering,
Here output undulator is 21 - cells long, and composed of 9 uniform plus 12 tapered cells. Grey lines refer to single shot realizations, the black line refers to the average over a hundred realizations.} \label{3Pz}
\end{figure}

\begin{figure}[tb]
\includegraphics[width=1.0\textwidth]{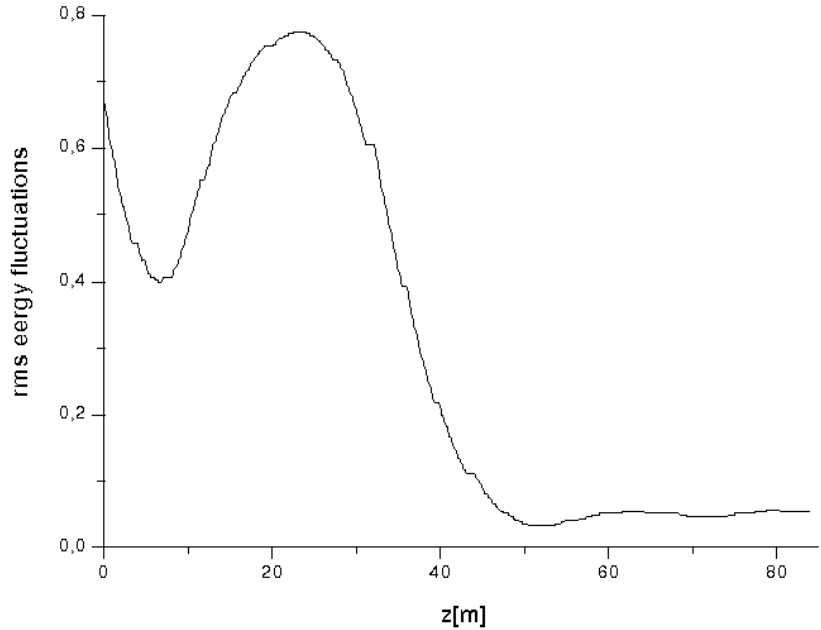}
\caption{rms energy deviation from the average as a function of the
distance inside the output undulator in the case of undulator tapering. Here the output undulator is 21 - cells long, and composed of 9 uniform plus 12 tapered cells.} \label{3rms}
\end{figure}

\begin{figure}[tb]
\includegraphics[width=1.0\textwidth]{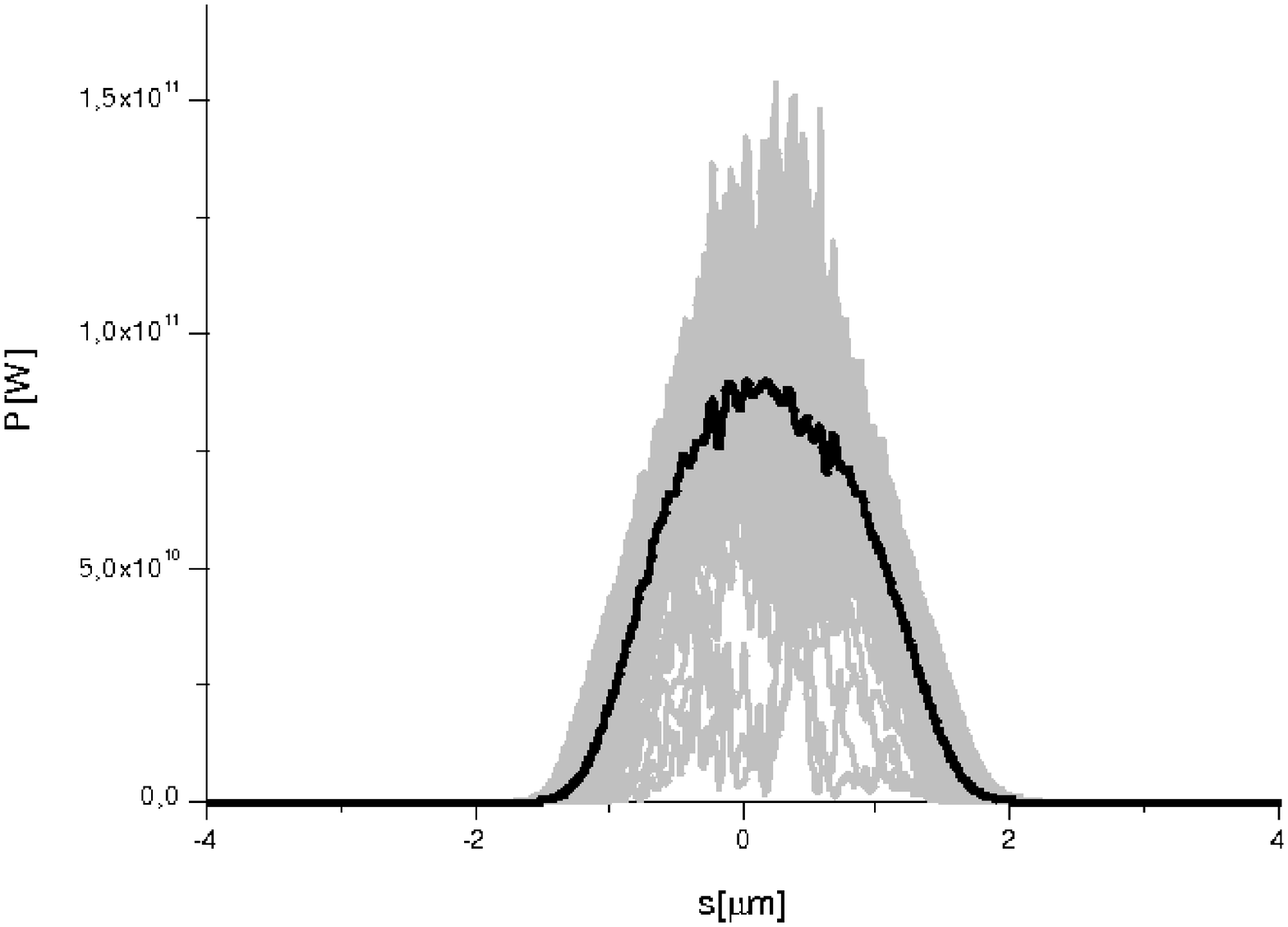}
\caption{Power distribution of the X-ray radiation pulse in the case
of undulator tapering. Here the output undulator is  21 - cells long and composed of 9 uniform and 12 tapered cells. Grey lines refer to single shot realizations, the black line refers to the average over a hundred realizations.} \label{3Pout}
\end{figure}

\begin{figure}[tb]
\includegraphics[width=1.0\textwidth]{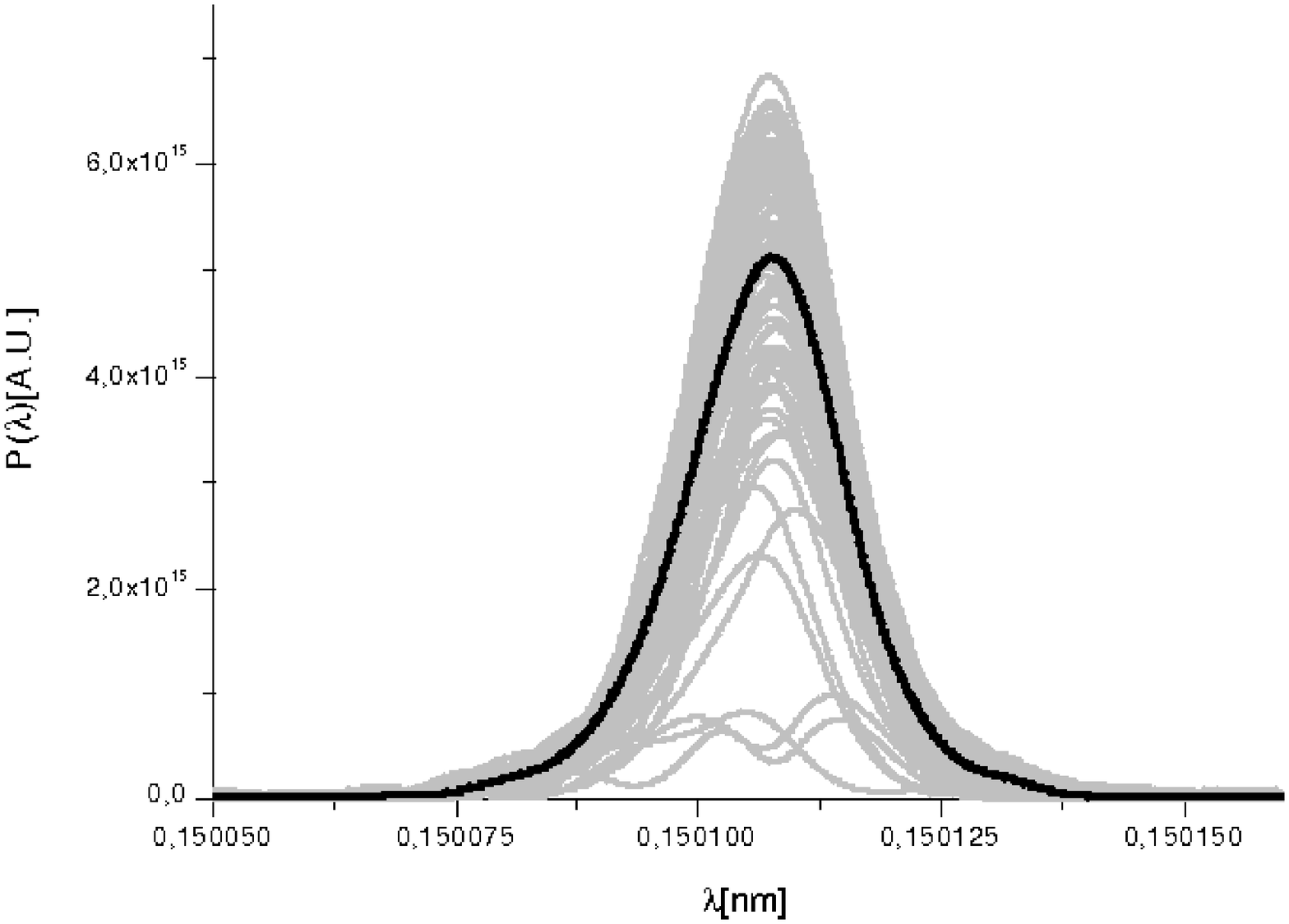}
\caption{Spectrum of the X-ray radiation pulse in the case of
undulator tapering. Here the output undulator is 21 - cells long and composed of 9 uniform and 12 tapered cells. Grey lines refer to single shot realizations, the black line refers to the average over a hundred realizations.} \label{3Spout}
\end{figure}
Fig. \ref{3Pz} shows the increase of energy in the radiation pulse as a function of the position inside the undulator. The rms energy deviation from the average as a function of the distance inside the output undulator is shown in Fig. \ref{3rms}. Fluctuations at the output of the device are less than $10\%$.  Fig.re \ref{3Pout} and Fig. \ref{3Spout} respectively show the output power and spectrum. In synthesis, the outstanding spectral properties (the bandwidth is still of order $10^{-4}$) and the good stability in terms of shot-to-shot energy fluctuations is retained, while the power is increased to the $100$ GW-level.

\section{\label{sec:7} Flexibility of the self-seeding setup installed in the LCLS baseline undulator}

\begin{figure}[tb]
\includegraphics[width=1.0\textwidth]{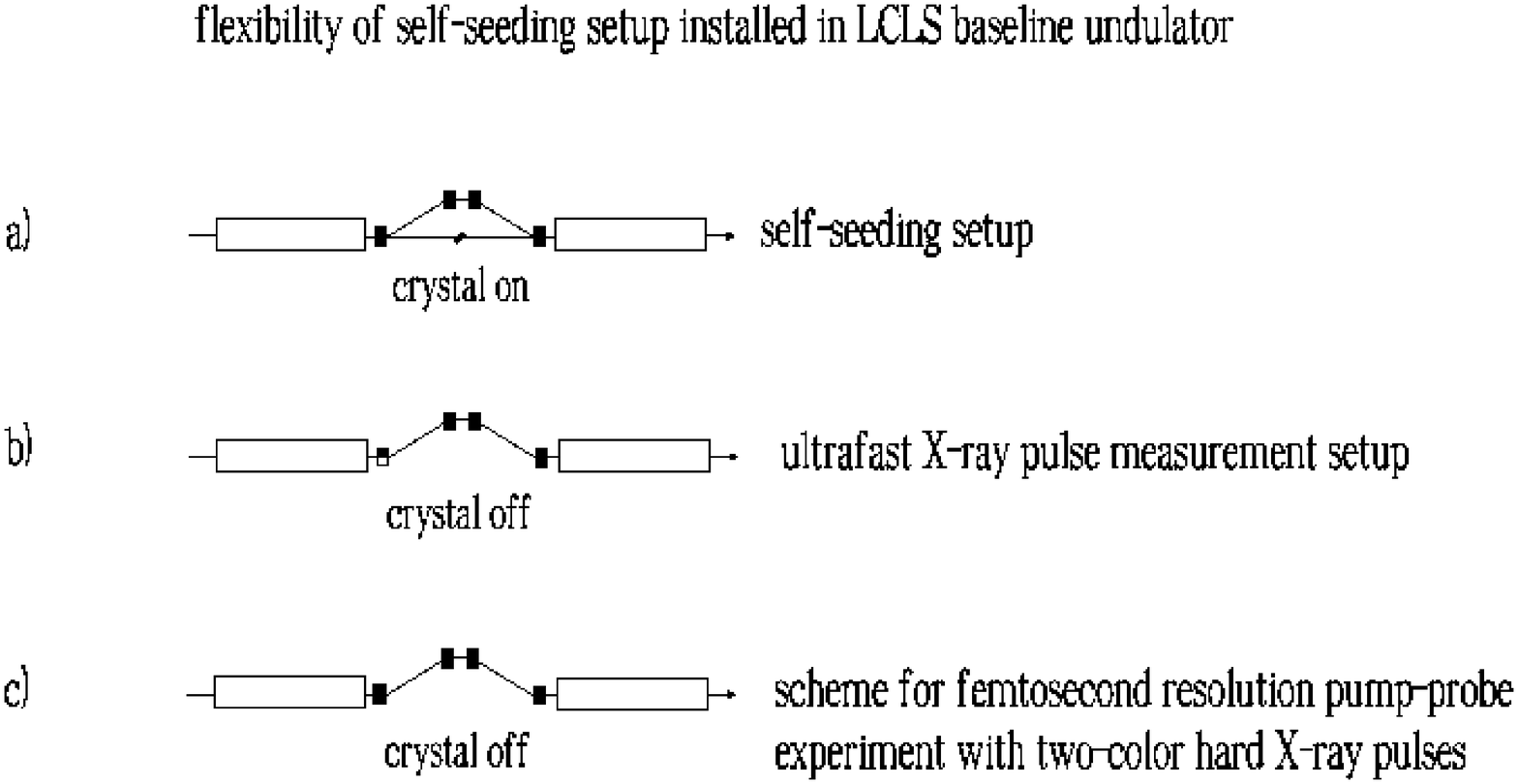}
\caption{Possible uses of the self-seeding setup, demonstrating the flexibility of our scheme. (a) self-seeding option; (b)  ultrafast X-ray pulse measurement option; and (c) pump-probe experiment option.} \label{lcls5}
\end{figure}

\begin{figure}[tb]
\includegraphics[width=1.0\textwidth]{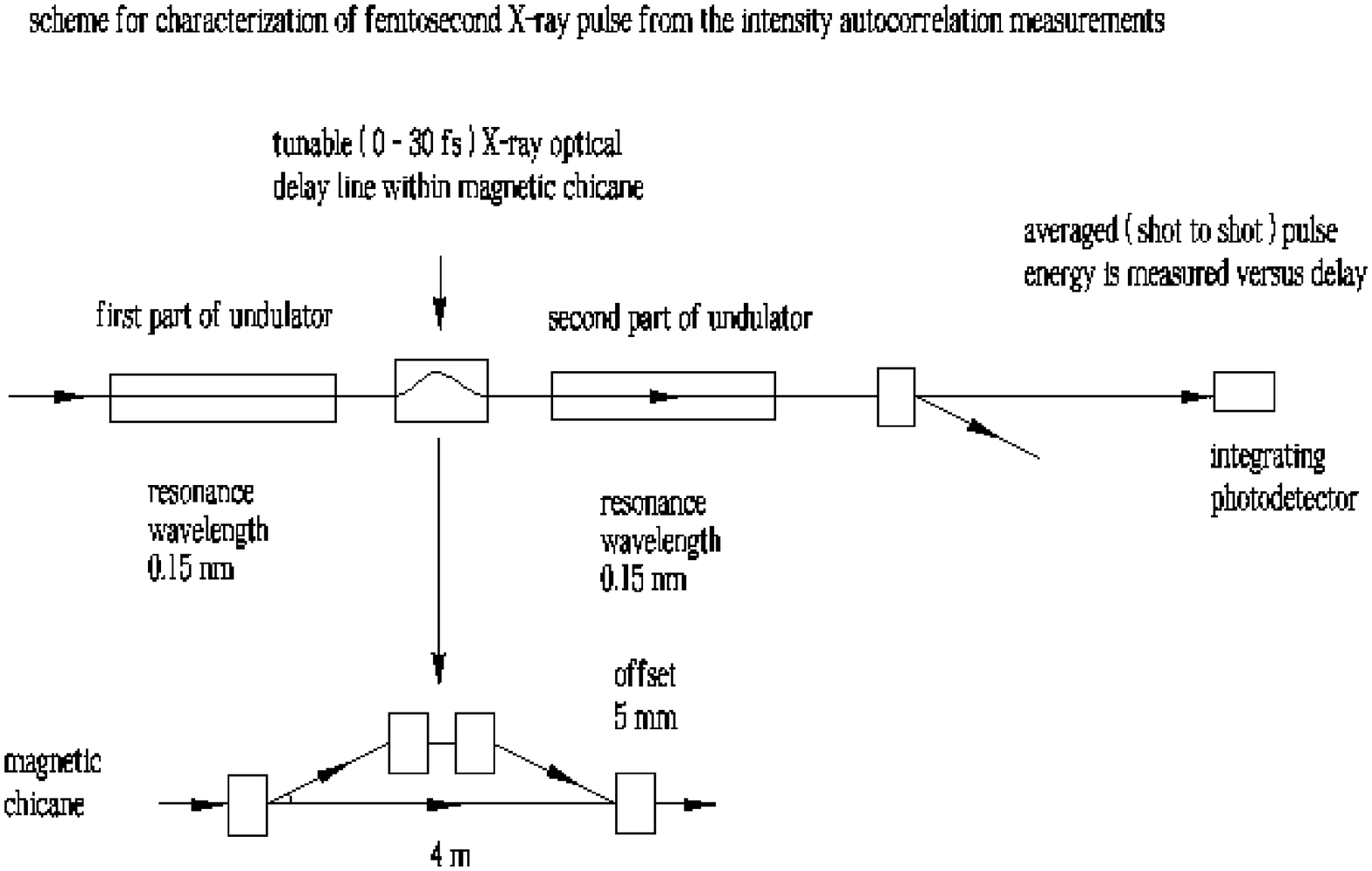}
\caption{Experimental layout for ultra-fast X-ray pulse measurement based on the use of the self-seeding setup hardware. Only $8$ out of $11$ cells of the first part of undulator are activated. Pulse and electron bunch are separated at the exit of the first part of the undulator. One is variably delayed with respect to the other. The modulation of the electron bunch is washed out. The x-ray SASE pulse and the "fresh" electron bunch are overlapped in the second part of undulator, where  only $8$ cells are activated, exactly as in the first part. The averaged X-ray pulse energy at the setup exit is measured versus the delay, yielding the autocorrelation trace.} \label{lcls7}
\end{figure}
\begin{figure}[tb]
\includegraphics[width=1.0\textwidth]{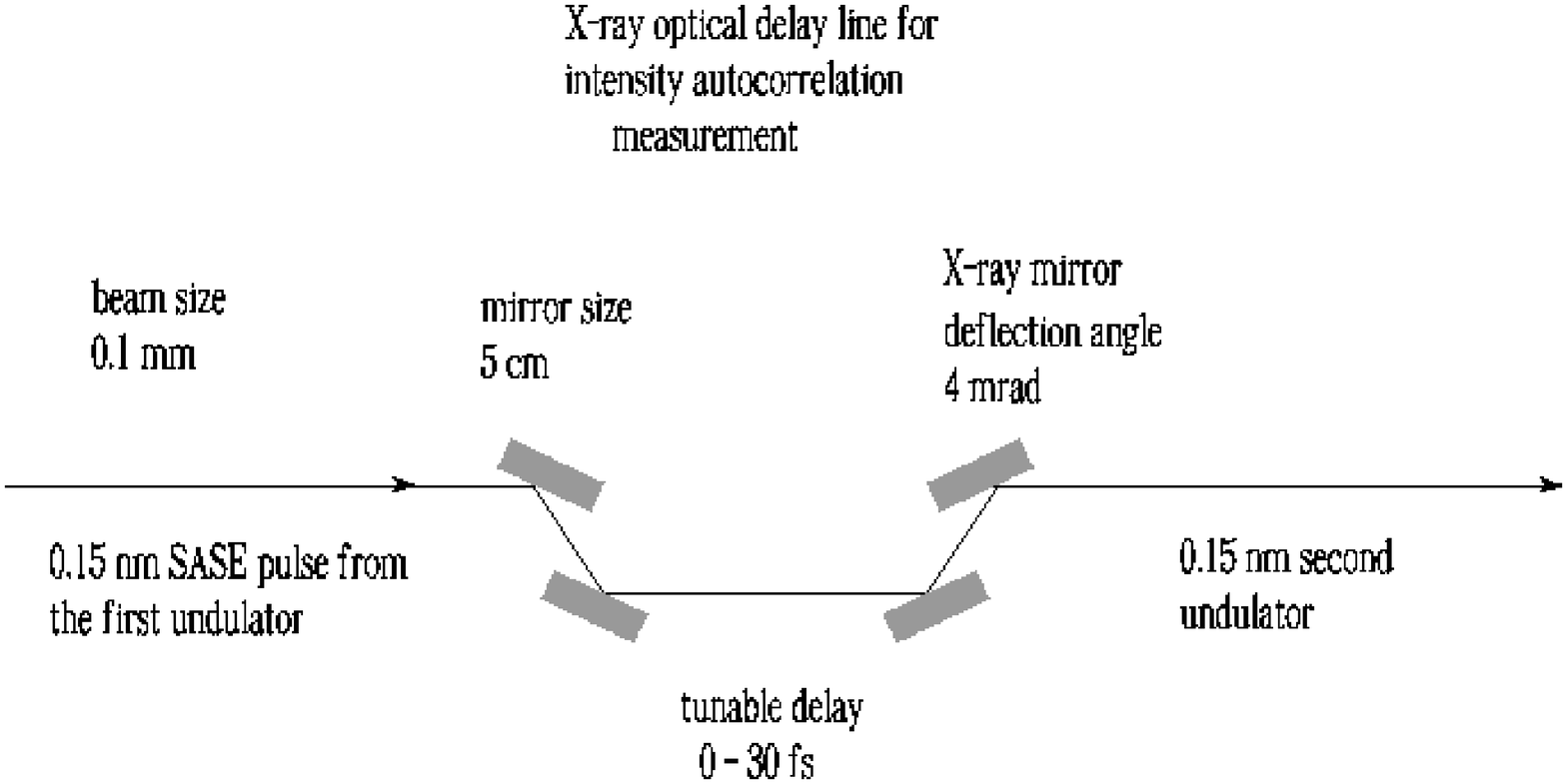}
\caption{X-ray optical system for delaying the SASE X-ray pulse with respect to the electron bunch. The X-ray optical system can be installed within the magnetic chicane of the self-seeding setup in placed of the single crystal monochromator.} \label{lcls8}
\end{figure}

The two-undulator configuration in Fig. \ref{lcls1} can be naturally taken advantage of in different schemes, as shown in Fig. \ref{lcls5}. The upper figure (a) shows the present self-seeding scheme.

The middle figure (b) refers to the ultra short X-ray pulse measurement setup proposed in \cite{OUR05}, which makes use of the magnetic chicane. The measurement of X-ray pulses on the
femtosecond time-scale constitutes an unresolved problem.
In fact, it is possible to create sub-ten femtosecond X-ray pulses from LCLS in the low-charge mode of operation, but not to measure their duration. In \cite{OUR05} we proposed a novel method for measuring the duration of femtosecond X-ray pulses from XFELs.  The method is based on the measurement of the autocorrelation function of the X-ray pulses. The setup in Fig. \ref{lcls5}, similar to that described in \cite{OUR05}, may be used at LCLS to this purpose, see Fig. \ref{lcls7}. The electron bunch enters the first part of the baseline undulator and produces SASE radiation with a ten MW-level power. After the first  part of the undulator the electron bunch is guided through the magnetic chicane, whose function is both to wash out the electron bunch modulation, and to create the necessary offset for the installation of an X-ray optical delay line consisting of a mirror chicane to delay the SASE radiation relatively to the bunch, Fig. \ref{lcls8}. The glancing angle of the X-ray mirrors is small as $2$ mrad. Inside the experimental hall, the transverse size of the radiation requires long mirrors, in the meter size. In contrast to this, at the undulator location the transverse size of the photon beam is smaller than $0.1$ mm, meaning that the mirror length would be just about $5$ cm. Moreover, the low-charge mode of operation relaxes heat-loading issues. After the chicane, electron beam and X-ray radiation pulse produced in the first part of the undulator enter the second part of the undulator, which is resonant at the same wavelength.  In the second part of the undulator the first X-ray pulse acts as a seed and overlaps with the lasing part of electron bunch. Therefore, the output power rapidly grows up to the GW-level. First and second part of the undulator are identical, i.e. they are composed of the same  number of active cells, seven, and operate in the linear FEL amplification regime. The relative delay between electron bunch and seed X-ray pulse can be varied by the X-ray optical delay line installed with the magnetic chicane.  The subsequent measurement procedure consists in recording the shot-to-shot averaged energy per pulse at the exit of the second part of undulator as a function of the relative delay between electron bunch and seed X-ray pulse with the help of an integrating photodetector. This yields the autocorrelation function of the X-ray pulse, see \cite{OUR05} for
more details.

Finally, the lower figure (c) in Fig. \ref{lcls5} shows the advantage of our two-undulator setup when dealing with pump-probe techniques. The production of ultrafast, high power, coherent X-ray pulses is of great importance when it comes to time-resolved experiments, which are used to monitor time-dependent phenomena. In typical pump-probe experiments, a short pump pulse is followed, after a given time, by a short probe pulse. Femtosecond pump-probe capabilities have been available for some years at visible wavelengths. Today there is a strong interest in extending these techniques to the X-ray region: in fact, they enable direct investigation of structural changes with atomic resolution. The synchronization between pump and probe pulses, which should have different wavelengths, remains one of the main technical problems for reaching this kind of pump-probe capabilities. Combination of high-power femtosecond optical pump with femtosecond
hard X-ray probe pulses provides a unique tool for the study of ultra-fast phenomena. Light-triggered, time-resolved studies with sub-10 fs resolution obviously require the same level of synchronization between optical and hard X-ray pulses.


\begin{figure}[tb]
\includegraphics[width=1.0\textwidth]{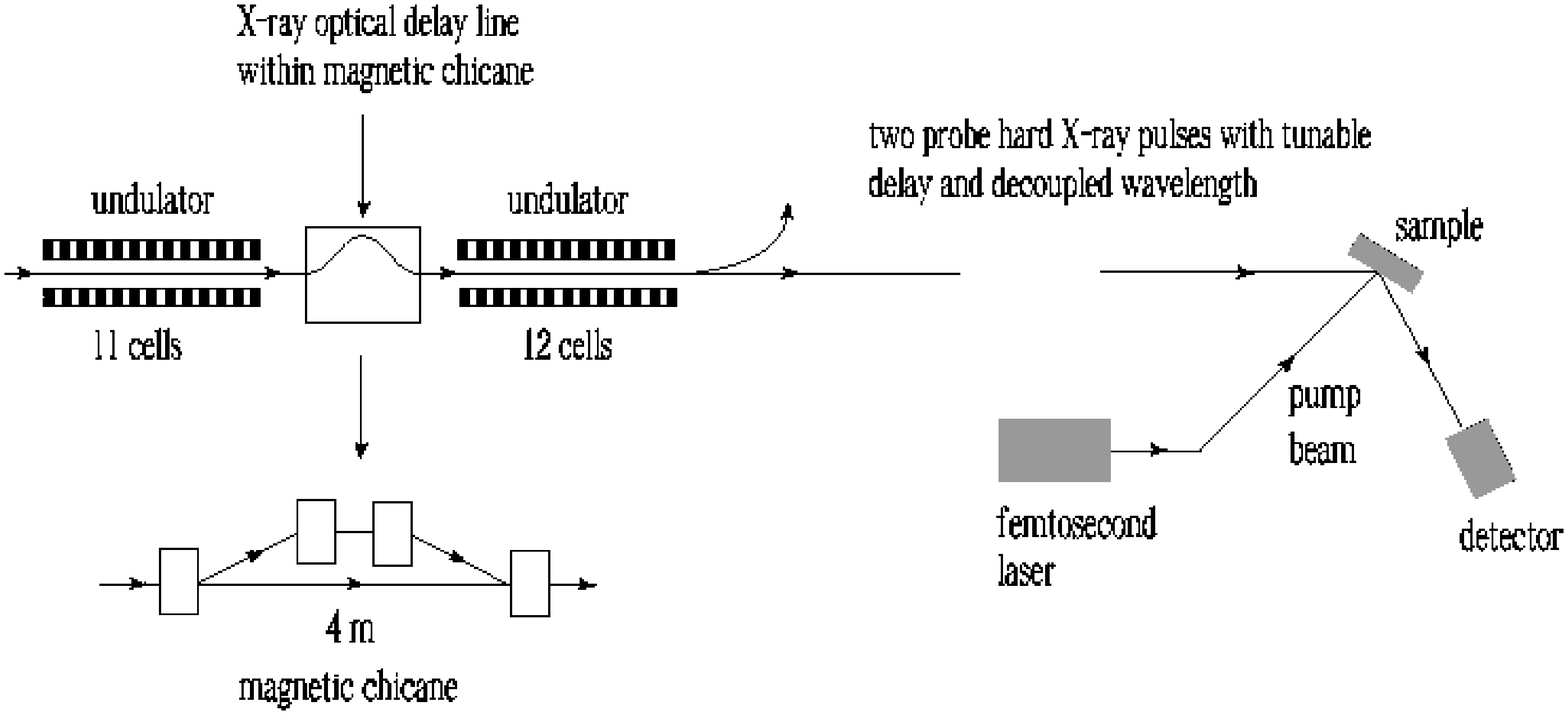}
\caption{Scheme for optical pump$/$hard X-ray probe experiments at LCLS  employing two hard X-ray pulses generated by the same electron bunch from two different parts of LCLS baseline undulator.} \label{lcls10}
\end{figure}

\begin{figure}[tb]
\includegraphics[width=1.0\textwidth]{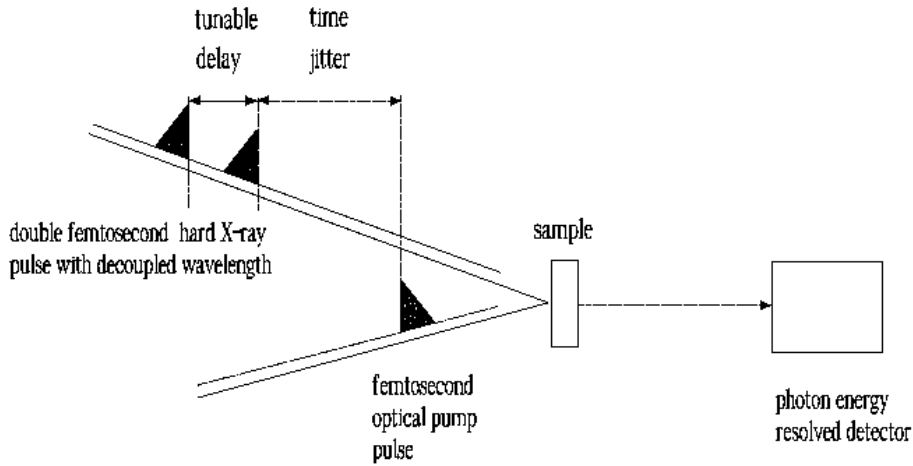}
\caption{A novel scheme for optical pump- hard X-ray probe- experiments. The principle of operation the scheme proposed for LCLS is essentially based on the exploitation of the time jitter between pump and probe pulses. We shift the attention from the problem of synchronization between pump and probe pulses to the problem of synchronization between two probe hard X-ray pulses with tunable delay and decoupled wavelengths. It is proposed to derive both probe hard X-ray pulses from the same electron bunch, but from different parts of the LCLS baseline undulator, which has a few-percent adjustable gap.  This eliminates the need  for synchronization
and cancels jitter problems.} \label{lcls9}
\end{figure}

Here we propose a method to get around this obstacle by using an optical pulse as pump and two hard X-ray pulses with tunable delay between them and decoupled wavelengths as probe\footnote{Obviously the proposed setup can also be used for hard X-ray pump $/$ hard X-ray probe experiments with decoupled wavelengths.}.  The overall scheme is shown in Fig. \ref{lcls10}. Two hard X-ray pulses generated by the same electron bunch from two different parts of the LCLS baseline undulator are delayed with the help of an optical delay line installed within the magnetic chicane. The pulses are tuned to a few percent different resonance wavelengths.  Delaying the two hard X-ray pulses with decoupled wavelengths may also be done in the experimental hall with dispersive X-ray optical elements (crystals), but in this case the intensity on the sample becomes weak. This is due to the fact that the SASE radiation has a rather broad bandwidth. As a crystal selects a much smaller bandwidth, most of the photons are discarded. Much larger throughput can be achieved in the situation when the delay is provided in the baseline undulator, since in this case the delay can be generated without dispersive X-ray optics.  The idea is to exploit the self-seeding setup, and to install an X-ray mirror chicane within  the magnetic chicane. The function of the mirror chicane is to delay the SASE radiation pulse generated in the first part of undulator relatively to the electron bunch and, therefore, also relatively to the second SASE radiation pulse generated downstream of the magnetic chicane, in the second part of the undulator.

The method exploits the jitter (about $100$ fs) between the pump optical pulse and the double hard X-ray probe pulse, and does not require synchronization, Fig. \ref{lcls9}. The technique consists in using two separate hard X-ray counters and record counts for each value of the delay between the two probe hard X-ray pulses. The detected photoevents are then correlated, and the sample dynamics is determined from that correlation, Fig. \ref{lcls10}. During the light-triggered structural changes, each of the two detectors registers a certain number of photocounts. Let us indicate these numbers with $K_1$ and $K_2$ respectively. We also assume that the shot-to-shot energy fluctuations are small, i.e. within a few percent, which is to be expected given the poor longitudinal coherence of the SASE pulses. $K_1$ and $K_2$ are further normalized to the incomining number of photons in the two radiation pulses. The incoming number of photons can be measured by two nondestructive detectors.

After each shot, the electronics multiplies $K_1$ by $K_2$, and passes the product to an averaging accumulator, where it is added to the previously stored sum of count products. Finally, the total sum is divided by a number of shots.  A tunable delay $\tau$ is introduced between the two probe hard X-ray pulses at the level of the LCLS undulator setup. This can be done with the same X-ray mirrors chicane to be used for the ultrafast X-ray pulse measurement considered above. Therefore, our setup using the self-seeding magnetic chicane for allowing the installation of an X-ray optical delay line can be used for two purposes: for ultrafast X-ray pulse measurements and for optical-pulse pump$/$double hard X-ray pulse probe experiments. The result of the above operations is the (shot to shot) averaged count product $\langle K_1\cdot K_2 \rangle$, $\langle ... \rangle$ indicating an average over an ensemble of shots. Different shots are characterized by a different jitter.  Let us call with $P(t)$ the probability of having a certain time delay $t$. Here time zero corresponds to the pump event, so that prior to the pump ($t<0$) one has no signal at all. Therefore  we can write

\begin{eqnarray}
\langle K_1\cdot K_2 \rangle = \int_{0}^{\infty} P(t) \cdot K_1(t) K_2(t+\tau) d t ~.
\label{K1K2}
\end{eqnarray}
The quantity in Eq. (\ref{K1K2}) obviously contains information about the signal function. Our method aims to a resolution of a few fs: in fact, we propose to apply it to the study phenomena with characteristic times in the order of a few tens fs. We therefore suppose that the jitter is significantly larger than the characteristic time of the phenomenon under study, at
least a few 100 fs. In this case we can assume that the shape of the probability function under the integral is nearly flat, and we can take it out of the integral sign. If we further assume that a percent difference in the energy of the incoming hard X-ray beam has no influence on the signal function\footnote{The two pulses have a difference in frequency in the order of a few percent in order to separately record the number of photocounts from the two probe pulses.}, we can also write $K_1(t) = K_2(t) = f(t)$, thus obtaining the signal autocorrelation function:

\begin{eqnarray}
\langle K_1\cdot K_2 \rangle = P(0) \cdot \int_{0}^{\infty}   f(t) f(t+\tau) d t = \mathcal{A}(\tau)~.
\label{Auto}
\end{eqnarray}
An alternative possibility of post processing of the collected
data consists in selecting events characterized by a given value $K_2$ when $K_1 = 0$. This is the case for $t<0$. Given a certain delay $\tau$, one has non-zero counts for $K_2$ and zero counts for $K_1$ for $-\tau <t< 0$, so that in this case the average sum of the counts $K_2$ equals to

\begin{eqnarray}
\langle  K_{2_{|K_1=0}} \rangle = P(0) \int_{-\tau}^{0}  f(t+\tau) d t ~,
\label{Auto2}
\end{eqnarray}
In this case, the signal function can be reconstructed from the derivative  of Eq. (\ref{Auto2}) with respect to $\tau$.

\begin{figure}[tb]
\includegraphics[width=1.0\textwidth]{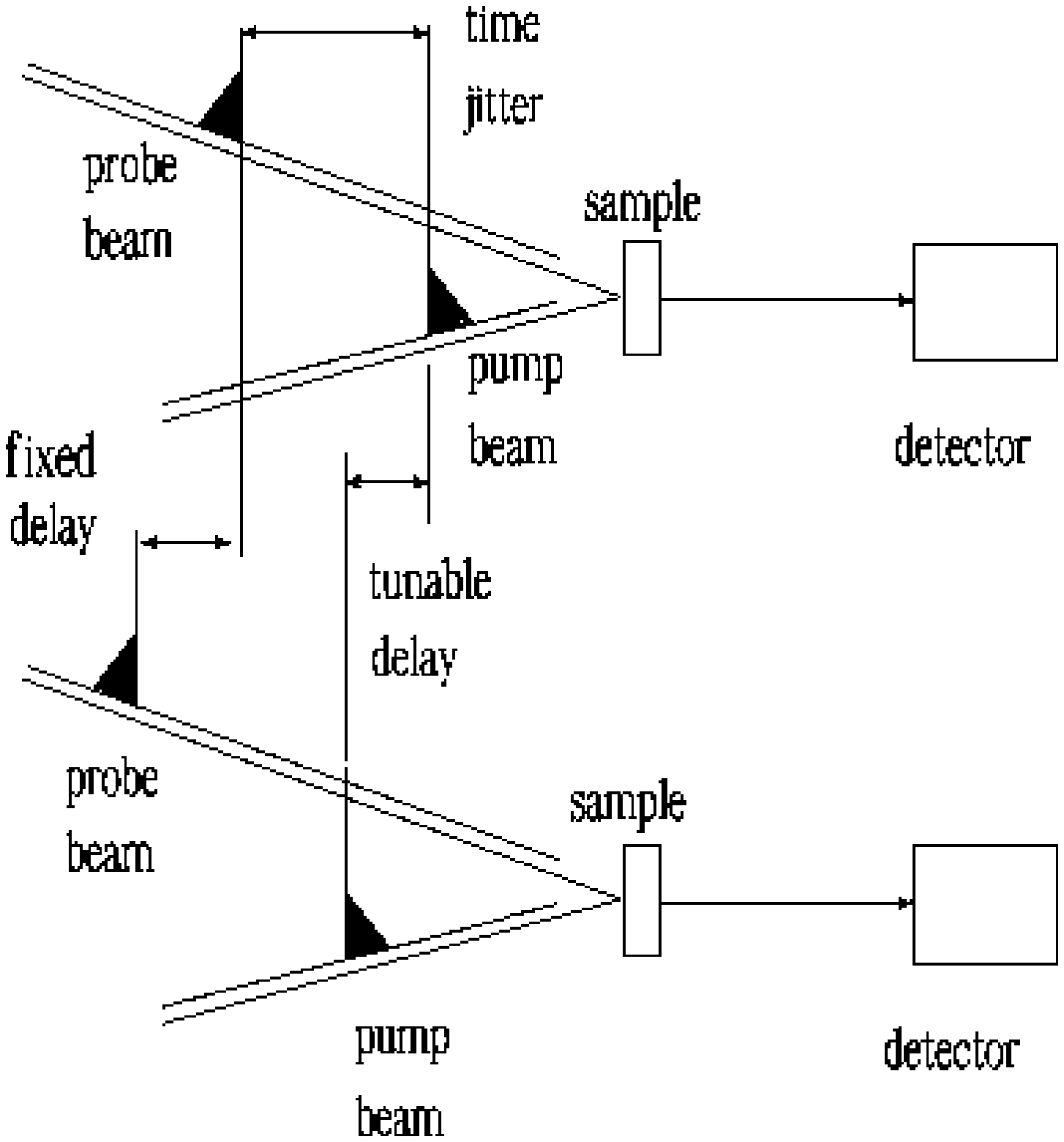}
\caption{A natural generalization of the pump-probe technique based on the correlation principle. In this case, the two SASE pulses are shifted in the transverse direction within the LCLS baseline undulator system, and impinge on two spatially separated samples. This spatial separation eliminates the need for decoupled wavelengths in order to separately record the number of counts from the two probe pulses. Possible applications will cover inelastic scattering measurements as well.} \label{lcls11}
\end{figure}
\begin{figure}[tb]
\includegraphics[width=1.0\textwidth]{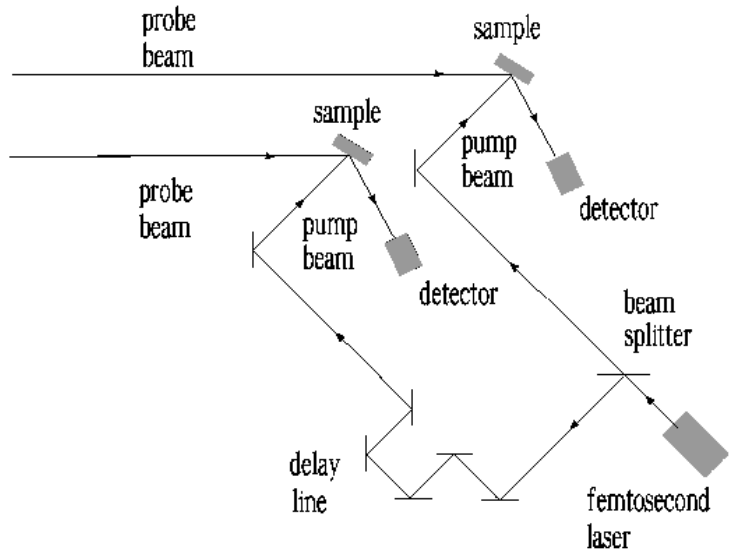}
\caption{Sketch of principle for the optical pump$/$hard X-ray probe
technique based on the use of time jitter between double pump and  double probe pulses. This technique exploits two spatially separated samples and records the correlation of the counts for each value of the delay between two pump-laser pulses.} \label{lcls12}
\end{figure}
\begin{figure}[tb]
\includegraphics[width=1.0\textwidth]{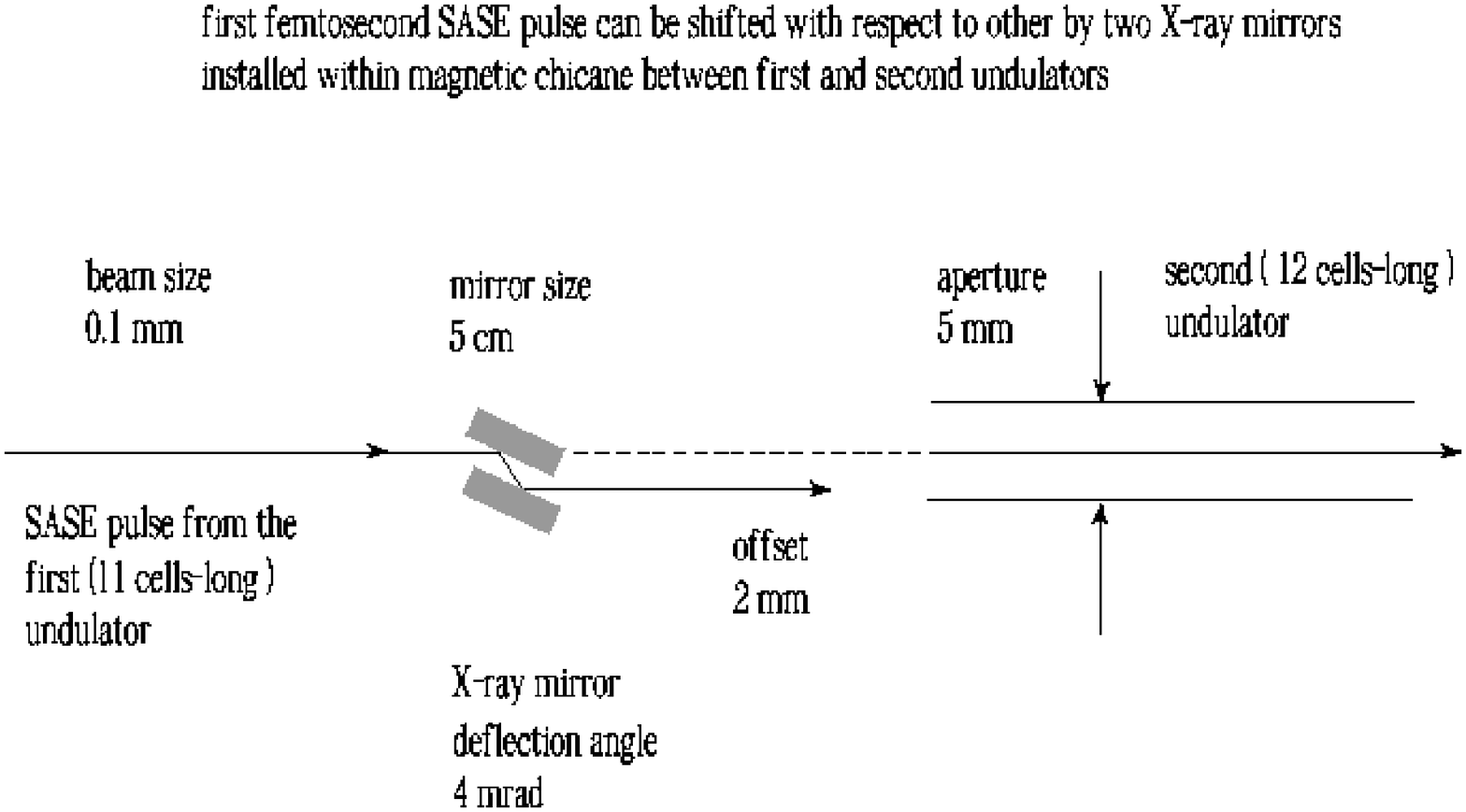}
\caption{Scheme for horizontally shifting the first femtosecond SASE
pulse with respect to the second within the LCLS baseline undulator system. Two X-ray mirrors can be installed within the self-seeding magnetic chicane between the first undulator (11-cells long) and the second (12-cells long).} \label{lcls13}
\end{figure}
\begin{figure}[tb]
\includegraphics[width=1.0\textwidth]{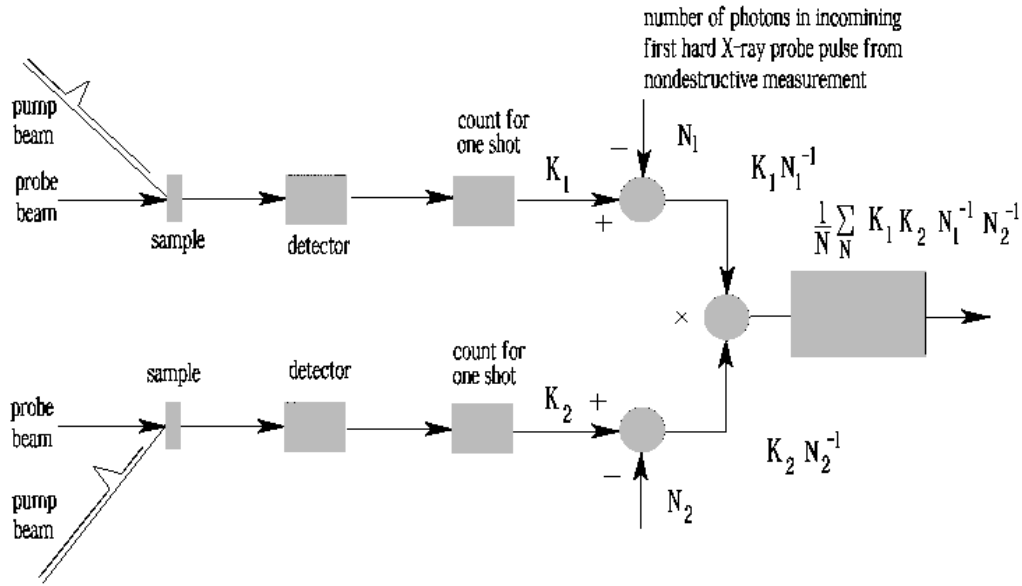}
\caption{Correlation apparatus for analysis of sample dynamics.} \label{lcls14}
\end{figure}
\begin{figure}[tb]
\includegraphics[width=1.0\textwidth]{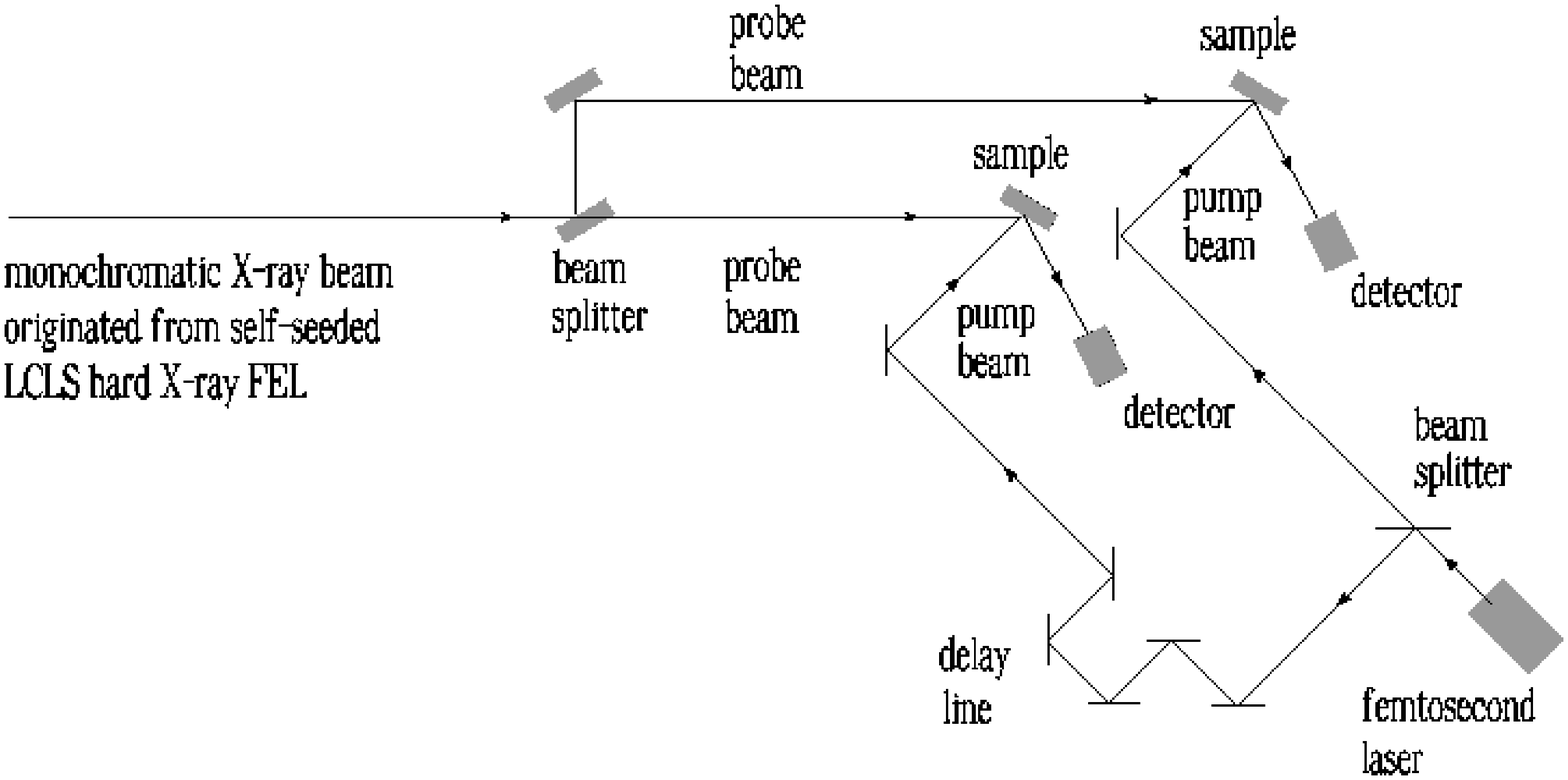}
\caption{Scheme for pump-probe experiments employing two probe pulses originating from the splitting of a single monochromatic  X-ray pulse produced by the self-seeded LCLS hard X-ray FEL in high-power mode of operation. Splitting the hard X-ray pulse may also be done in SASE mode of operation without monochromatization, but in this case the intensity on the sample is weaker. This is due to the fact that the SASE radiation has a rather broad bandwidth. Since a crystal-splitter selects a much smaller bandwidth, most of the photons in this situation are discarded.} \label{lcls15}
\end{figure}
In some experimental situations, the scheme for pump-probe experiments depicted in Fig. \ref{lcls9} is not optimal.  A possible technical extension (see Fig. \ref{lcls11} and \ref{lcls12}) is to use two spatially separated sets of samples (see Fig. \ref{lcls13}), pump and probe pulses and record the correlation of the counts for each value of the delay $\tau$ between two pump pulses . The detected photoevents are than correlated and the sample dynamics is determined from that correlation. The version of the apparatus for measuring the average count product is depicted in Fig. \ref{lcls14}. This new approach offers a number of significant advantages. In the scheme in Fig. \ref{lcls9}, the wavelength difference between the two pulses cannot be more than a few percent in LCLS baseline case, and this constitutes a significant restriction for the applicability of the method. The spatial separation eliminates the need for different wavelengths in order to separately record the number of counts from the two probe pulses. As a result, possible applications of the scheme in Fig. \ref{lcls12} also cover inelastic scattering experiments.

The new  experimental setup is sketched in Fig. \ref{lcls11} and Fig. \ref{lcls12}.  The pump-laser pulse is splitted, delayed and incident on two spatially separated samples. The two SASE pulses, generated by the same electron bunch from two different parts of the baseline undulator, are shifted in transverse direction with the help of X-ray mirrors. The idea is sketched in Fig. \ref{lcls13}. The two SASE pulses can be horizontally separated by two mirrors installed within the self-seeding magnetic chicane after the first part of undulator (11-cells long). The horizontal offset can be about $2$ mm. Additionally, the mirrors can also be used to generate a few microradians deflection-angle, which is not important within the undulator, but will create further separation of a few millimeters at the position of the experimental station.

A second possible experimental setup with spatially separated samples is based on the use of an X-ray splitting setup installed in experimental hall (see Fig. \ref{lcls15}). High throughput can be achieved when the bandwidth of the X-ray beam is sufficiently small. This scheme makes use of the monochromatic X-ray beam originated from the self-seeded LCLS hard X-ray FEL.

\begin{figure}[tb]
\includegraphics[width=1.0\textwidth]{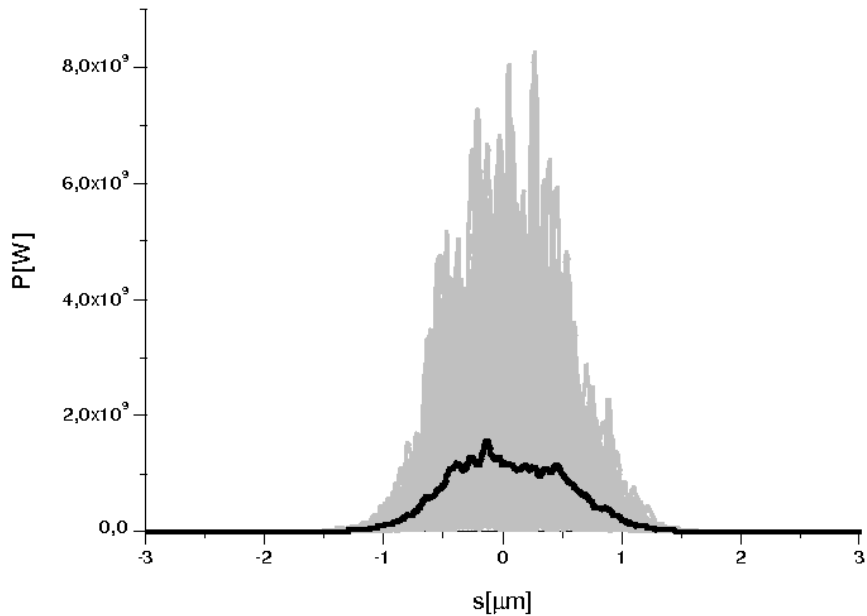}
\caption{Pump-prob experimental setup. Power distribution of the
second X-ray radiation pulse.  Here the second undulator  is12 - cells long and resonant at 0.1527 nm wavelength. Grey lines refer to single shot realizations, the black line refers to the average over a hundred realizations.} \label{PP_Pout}
\end{figure}

\begin{figure}[tb]
\includegraphics[width=1.0\textwidth]{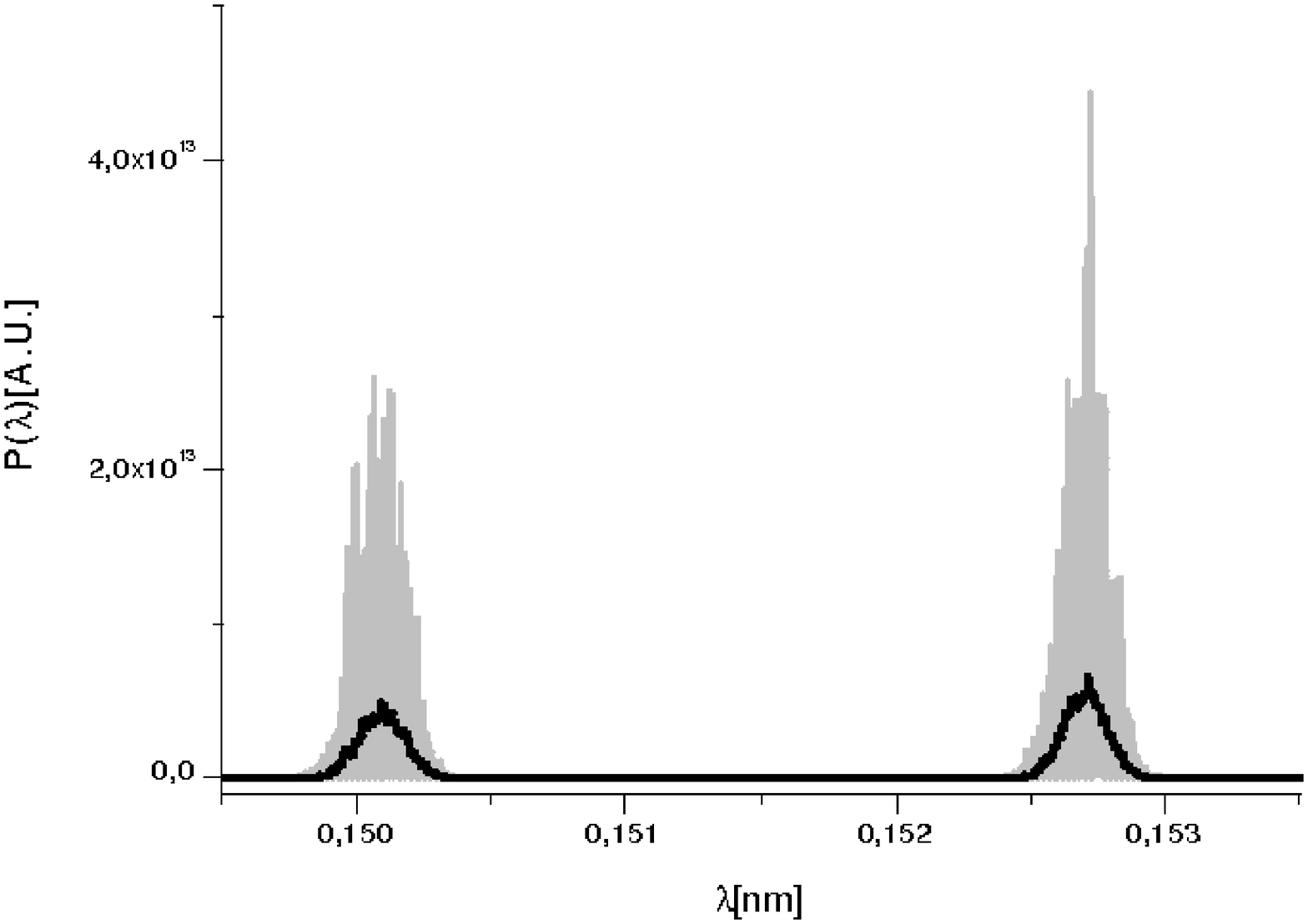}
\caption{Pump-probe experimental setup. Spectrum of  double hard X-ray probe pulse. Here the first undulator is11- cells long and resonant at 0.1500 nm wavelength. The second undulator is 12- cells long and resonant at 0.1527 nm wavelength. Grey lines refer to single shot realizations, the black line refers to the average over a hundred realizations. The two colors are clearly visible.} \label{PP_SPout}
\end{figure}

In the previous Section we already performed simulations in the SASE mode for the first 11 cells of the LCLS undulator. The power distribution is shown in Fig. \ref{1Pin}. The simulation for the second undulator part consists once more of SASE simulations, where the electron beam is modeled using the values of energy loss and energy spread at the end of the first part. The resonant wavelength is now set, for exemplification purposes, to about $\lambda_1 = 0.1501$ nm and $\lambda_2 = 0.1527$ nm. The output power distribution and spectrum after 12 cells are shown respectively in Fig. \ref{PP_Pout} and Fig. \ref{PP_SPout}. A shot-to-shot averaged peak power of about $1.5$ GW is foreseen\footnote{If needed, one may increase the output power of the two-color hard X-ray pulses up to ten GW level by application fresh bunch technique. For this, one needs to install an additional magnetic chicane and to use the full number of undulator cells. See our previous paper \cite{OUR01} for more details.}.


\section{\label{sec:8} Conclusions}

In this paper we demonstrated that monochromatization of x-ray pulses from XFELs is of great importance in the LCLS upgrade program. In fact, aside for an improvement of brilliance up to two orders of magnitude, a highly monochromatized pulse opens up many other ways to enhance the capacity of the LCLS.

From this standpoint, we demonstrated that monochromatization effectively allows one to use an undulator tapering technique enabling a high-power mode of operation. Monochromatization allows one to use X-ray phase retarders in the experimental hall too, leading with almost no additional costs to full polarization control at the LCLS hard X-ray FEL. In this paper we have also shown how monochromatization enables the use of crystal deflectors in the LCLS X-ray transport system, thus providing the possibility for building new experimental halls (surface structure) for experiments in the hard X-ray wavelength range.

Many other applications of monochromatization for enhancing LCLS
capacities  may be found. However, in order to keep this paper within reasonable size we did not discuss them here, and we will address them in future publications.

The progress which can be achieved with the methods considered here is based on the novel self-seeding scheme with single crystal monochromators \cite{OURX,OURY2}, which is extremely compact and can be straightforwardly realized in the LCLS baseline undulator. The proposed self-seeding setup takes almost no cost and
time to be implemented at the LCLS. The LCLS baseline remains the best place for testing our new self-seeding scheme. In its turn, this will allow for a very fast, significant enhance of the LCLS baseline capacity.

\section{Acknowledgements}

We are grateful to Massimo Altarelli, Reinhard Brinkmann, Serguei
Molodtsov for their support and their interest during the compilation of this work. We thank Edgar Weckert for many useful discussions.

\end{document}